\documentclass[prb,twocolumn,preprintnumbers,superscriptaddress,showpacs]{revtex4-1}
\usepackage{graphicx}
\usepackage{amsmath}
\usepackage{amssymb}
\usepackage{comment}
\usepackage[colorlinks,pdfstartview=FitH]{hyperref}
\hypersetup{linkcolor=blue,citecolor=blue,filecolor=black,urlcolor=blue}

\renewcommand*{\l}{\ensuremath{{\mathbf{l}}}}
\renewcommand*{\v}{\ensuremath{{\mathbf{v}}}}
\newcommand*{\p}{\ensuremath{{\mathbf{p}}}}
\renewcommand*{\d}{\ensuremath{{\mathrm{d}}}}
\renewcommand*{\i}{\ensuremath{{\mathrm{i}}}}
\renewcommand*{\mod}{\ensuremath{ \ {\mathrm{mod} \ }}}

\makeatletter
\newcommand*\wt[1]{\mathpalette\wthelper{#1}}
\newcommand*\wthelper[2]{%
    \hbox{\dimen@\accentfontxheight#1%
        \accentfontxheight#11.15\dimen@
        $\m@th#1\widetilde{#2}$%
        \accentfontxheight#1\dimen@
    }%
}
\newcommand*\accentfontxheight[1]{%
    \fontdimen5\ifx#1\displaystyle
        \textfont
    \else\ifx#1\textstyle
        \textfont
    \else\ifx#1\scriptstyle
        \scriptfont
    \else
        \scriptscriptfont
    \fi\fi\fi3
}
\makeatother

\begin{document}

\author{Zhaoyu Han}
\affiliation{Department of Physics, Stanford University, Stanford, CA 94305, USA}
\author{Jing-Yuan Chen}
\affiliation{Institute for Advanced Study, Tsinghua University, Beijing, 100084, China}

\title{Solvable Lattice Hamiltonians with Fractional Hall Conductivity}
\date{\today}

\begin{abstract}
We construct a class of lattice Hamiltonians that exhibit fractional Hall conductivity. These Hamiltonians, while not being exactly solvable, can be controllably solved in their low energy sectors, through a combination of perturbative and exact techniques. Our construction demonstrates a systematic way to circumvent the Kapustin-Fidkowski no-go theorem and is generalizable.
\end{abstract}

\maketitle


\section{Introduction}
\label{sect_intro}

The topological states of matter has been an important theme in condensed matter physics since the discovery of the quantum Hall effect four decades ago \cite{klitzing1980new, tsui1982two, Laughlin:1983fy}. Over time as people's knowledge in this field broadened and understandings deepened, the construction of exactly solvable lattice models emerged as a promising approach to study a large variety of topological states as well as their interplays with global symmetries \cite{Kitaev:1997wr, Levin:2004mi, Chen:2011pg}. The success of the exactly solvable lattice model approach is at least three-fold: 1) these models provide unambiguous microscopic completions for the associated topological phases, many of which hypothesized on theoretical grounds, and demonstrate that they can be realized in solid state systems at least in principle; 2) all the interesting topological properties can be exactly solved for and thereby apprehended in an explicit manner; 3) these models are constructed not out of fortuity but out of systematic considerations under certain principles, and the physical and mathematical origin of those principles is profound \cite{Levin:2004mi, Chen:2011pg, Turaev:1992hq, Kirillov:2011mk, Kitaev:2011dxc}.

In condensed matter physics, one most important global symmetry a topological state may interplay (``be enriched'') with is the electromagnetic $U(1)$. Indeed, the quantum Hall effect, whose very advent opened the entire field, was primarily characterized by the integrally or fractionally quantized Hall conductivity, featuring such an interplay. A rather curious situation occurs, however, if we attempt to construct exactly solvable lattice Hamiltonians, applying the usual ingredients and wisdom that have generated so many successes, for the topological states that exhibit non-trivial Hall conductivity -- Such efforts never came to fruition. The fundamental reason behind this apparently disappointing situation was finally addressed by Kapustin and Fidkowski, stated by them in a no-go theorem \cite{kapustin2020local}: \emph{A non-trivial Hall conductivity is impossible in a gapped local commuting projector Hamiltonian with finite-dimensional local Hilbert space.}

The purpose of this work is to go beyond the usual approach of exactly solvable lattice models, and in particular beyond the limiting conditions stated in the Kapustin-Fidkowski no-go theorem, in order to construct lattice Hamiltonians that can be solved, at least in the low energy sectors, to find non-trivial Hall conductivity.

In Ref.~\cite{Chen:2019mjw} one of us studied the electromagnetic $U(1)$ enrichment for a large class of abelian topological phases (those which admit gapped boundary conditions were the electromagnetic $U(1)$ absent \cite{Kapustin:2010hk, Lin:2014aca}), including the cases with Hall conductivity, and found exactly solvable Lagrangians for them on \emph{effective} spacetime lattices (i.e. coarse grained spacetime manifolds). There it was also explained how and why, if we proceed with the usual wisdom \cite{Levin:2004mi, Kirillov:2011mk} in attempt to reduce an exactly solvable Lagrangian on a coarse grained spacetime to an exactly solvable toy model Hamiltonian on an actual spatial lattice, we would run into problems when the Hall conductivity is non-trivial. The occurrence of the problems is an embodiment of the Kapustin-Fidkowski no-go theorem. 

Through the study in Ref.~\cite{Chen:2019mjw}, however, a path was sketched towards constructing controllably solvable -- despite not being exactly solvable -- lattice Hamiltonians with non-trivial Hall conductivity. In this paper we elaborate on the simplest such cases and show that the Hamiltonians can indeed be solved, exhibiting the desired values of Hall conductivity which are fractional in general. The solution is obtained by a combination of perturbation theory and the exact techniques familiar in solving local commuting projector Hamiltonians. (We remark that a well-known example of Hamiltonian solved by a combination of perturbative and exact techniques is the celebrated Kitaev honeycomb model \cite{Kitaev:2006lla} which is a non-abelian spin liquid, although the detailed procedure is very different from our present work.) Our work is conceptually straightforward and can be generalized to a larger class \cite{Chen:2019mjw} of topological phases coupled to electromagnetic $U(1)$, including the fermionic phases; we will elaborate on those generalizations in subsequent works.

This paper is organized as the following. In Section \ref{sect_Hamiltonian} we motivate the lattice Hamiltonians following the idea outlined in \cite{Chen:2019mjw}; we also explain the relation between our present work and the previous literature, in particular Refs.~\cite{geraedts2013exact,geraedts2017lattice} and \cite{DeMarco:2021erp}. In Section \ref{sect_LLES} we use perturbation theory to obtain the low energy subspace within each local Hilbert space. In Section \ref{sect_solution} we solve for the many-body ground state(s) and anyon excitation states using steps similar to those in solving local commuting projector Hamiltonians. In Section \ref{sect_Hall} we compute the Hall conductivity, and show it is precisely quantized to the values that we designated, despite that we have apparently used perturbation theory in the derivation. In Section \ref{sect_conclusion} we make concluding remarks.

\section{Hamiltonian}
\label{sect_Hamiltonian}

In this paper we focus on the simplest cases among the bosonic topological orders enriched by electromagnetic $U(1)$ considered in \cite{Chen:2019mjw}. In the continuum they are described by a class of doubled Chern-Simons (BF) theories
\begin{align}
S = \int d^3 x \: \left[ \frac{n}{2\pi} a d b - \frac{q}{2\pi} Ad a - \frac{p}{2\pi} A d b \right]
\label{doubleCS}
\end{align}
with $n, p, q$ being integers. Here $a, b$ are dynamical $U(1)$ gauge fields and $A$ is the electromagnetic $U(1)$ background. When the coupling to $A$ is absent, the theory can be reduced to the $\mathbb{Z}_n$ generalization of the toric code (see below), therefore $n$ determines the intrinsic topological order, with $n=1$ topologically trivial \cite{Witten:2003ya}. On the other hand, $p, q$ determine the $U(1)$ global symmetry enrichment. (In principle, $a, b$ can couple to two different $U(1)$ global symmetry backgrounds $A, B$, but in this work we will identify $B=A$.) This model can be viewed as certain ``double layered'' bosonic quantum Hall. Its Hall conductivity is $-2pq/n$, and it supports two types of anyons, one coupled to $a$ and the other to $b$, that have trivial self-statistics and $-2\pi/n$ mutual statistics, and carry electric charges $p/n$ and $q/n$ respectively \cite{Wen:1995qn}.

In \cite{Chen:2019mjw} it was shown that such topological orders along with the electromagnetic $U(1)$ enrichment admit effective Lagrangians on spacetime lattice, that retain all the formal properties and are exactly solvable. By viewing one direction of the spacetime lattice as the discretized time, through standard procedure, a spacetime lattice Lagrangian gives rise to a set of commutation relations, as well as a spatial lattice Hamiltonian were it in the usual cases; in our case, however, related to the fact the Lagrangian is exactly solvable, instead of a Hamiltonian it turns out we actually get a set of strict constraints on the Hilbert space. While a set of constraints on the Hilbert space is not what we hope for, we can use them to motivate a lattice Hamiltonian, whose low energy sector reproduces and therefore replaces those constraints. The Hamiltonian thereby motivated may not be exactly solvable any more, but as long as its low energy sector can be controllably solved for, we have achieved our goal. The motivating procedure is reviewed in details in Appendix \ref{app_L_to_H}. Here we will present the results: the lattice Hilbert space, the commutation relations, and the lattice Hamiltonians for general integers $n, p, q$. 

\begin{figure}[t]
\includegraphics[width=.45\textwidth]{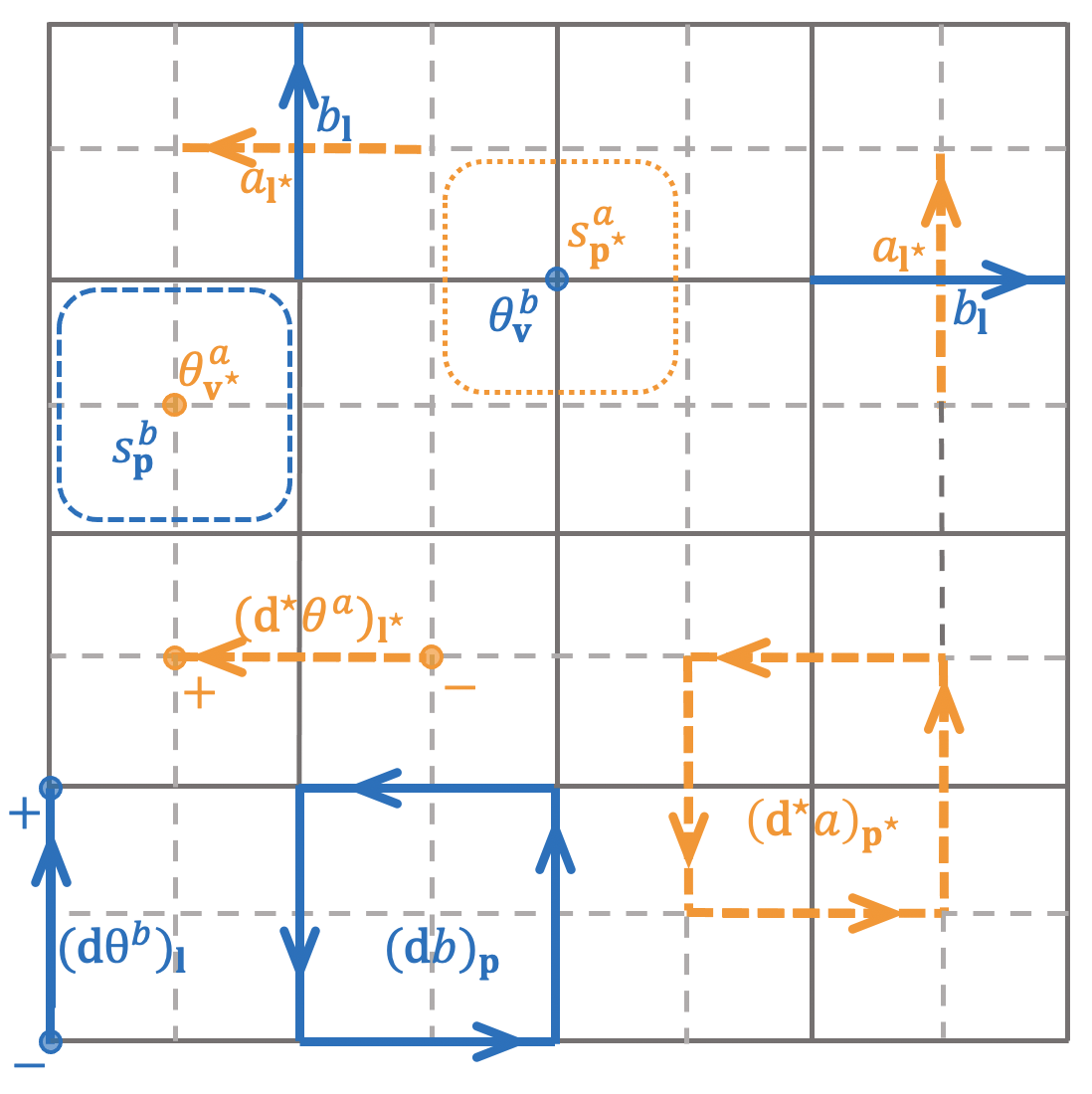}
\caption{Illustration for the lattice (solid line) and the dual lattice (dash line), the conjugate pairs of operators, and the lattice and dual lattice ``exterior derivatives''. Note we have picked the natural directions of the lattice links to be $+\hat{\mathbf{x}}$, $+\hat{\mathbf{y}}$, and the natural directions of the dual lattice links to be $90$-degrees counter-clockwise to the associated lattice link directions. When the direction label on a link (or a dual link) reverses, the associated operator picks a negative sign.}
\label{fig_lattice}
\end{figure}

We suppose the spatial lattice is a square lattice for simplicity (we can generalize all of our results to a triangulation with branching structure of an arbitrary two-dimensional spatial manifold). The local Hilbert spaces and their endowed operators are (see Fig.~\ref{fig_lattice} for illustration):
\begin{itemize}
\item
On each link $\l$, which is equivalent to a dual lattice link $\l^\star$ (with direction 90-degrees counter-clockwise to the direction of $\l$), there is a local Hilbert space endowed with a conjugate pair of real valued operators satisfying
\begin{align}
[b_\l, a_{\l^\star}] = \i \frac{2\pi}{n}.
\label{ba_comm}
\end{align}
\item
On each vertex $\v$, which is equivalent to a dual lattice plaquette $\p^\star$, there is a local Hilbert space endowed with a conjugate pair of integer/$U(1)$ valued operators satisfying
\begin{align}
[s^a_{\p^\star}, e^{\i\theta^b_\v}] = e^{\i\theta^b_\v}.
\label{sathetab_comm}
\end{align}
\item
On each plaquette $\p$, which is equivalent to a dual lattice vertex $\v^\star$, there is a local Hilbert space endowed with a conjugate pair of integer/$U(1)$ valued operators satisfying
\begin{align}
[s^b_\p, e^{\i\theta^a_{\v^\star}}] = e^{\i\theta^a_{\v^\star}}.
\label{sbthetaa_comm}
\end{align}
\end{itemize}
To have an intuitive picture in mind (which will appear to be an emergent picture that is not exact, after we introduce our Hamiltonian later), one may think of $a, b$ as dynamical $\mathbb{R}$ gauge fields on the lattice, such that they are effectively reduced to dynamical $U(1)$ gauge fields upon the introduction of the associated dynamical Dirac string variables $s^a$ and $s^b$. The ``reduction to $U(1)$'' can be understood in the following sense. The effective $U(1)$ fluxes associated with the dynamical gauge fields $a, b$ are
\begin{align}
f^b_\p \equiv (\d b - 2\pi s^b)_\p, \ \ \ \ \ f^a_{\p^\star} \equiv (\d^\star a - 2\pi s^a)_{\p^\star}
\label{generator_f}
\end{align}
where the lattice curl $\d b$ and the dual lattice curl $\d^\star a$ are illustrated in Fig.~\ref{fig_lattice}. Thanks to the Dirac string variables, if we sum $f^b_\p$ or $f^a_{\p^\star}$ over all the plaquettes $\p$ or $\p^\star$ on a closed space (consider a square lattice plane with periodic boundary conditions, forming a torus), we can have arbitrary $2\pi\mathbb{Z}$ values, reproducing the Dirac quantization condition for $U(1)$ gauge fluxes. Moreover, we can see that the $f^a, f^b$ fluxes are invariant under 
\begin{align}
& b_\l \: \rightarrow \: b_\l + 2\pi z^b_\l, \ \ \ \ \ s^b_\p \: \rightarrow \: s^b_\p + (\d z^b)_\p, \nonumber \\[.1cm]
& a_{\l^\star} \: \rightarrow \: a_{\l^\star} + 2\pi z^a_{\l^\star}, \ \ \ \ \ s^a_{\p^\star} \: \rightarrow \: s^a_{\p^\star} + (\d^\star z^a)_{\p^\star}
\label{1-form_Z}
\end{align}
where $z^b_\l$, $z^a_{\l^\star}$ are arbitrary $\mathbb{Z}$ valued transformations on the links and dual links respectively. Such invariances manifest the fact that $a, b$ are effectively reduced from $\mathbb{R}$ to $\mathbb{R}/2\pi\mathbb{Z} = U(1)$ gauge fields; they are known as $1$-form $\mathbb{Z}$ gauge invariances, see \cite{Chen:2019mjw} and Appendix \ref{app_L_to_H} (a general introduction to higher form symmetries can be found in \cite{Gaiotto:2014kfa}). Let us define the generators
\begin{align}
g^a_{\l^\star} \equiv e^{\i(\d^\star\theta^a-na)_{\l^\star}}, \ \ \ \ \ g^b_{\l} \equiv e^{\i(\d\theta^b-nb)_{\l}},
\label{generator_g}
\end{align}
where the notions of $\d\theta^b$ and $\d^\star\theta^a$ are illustrated in Fig.~\ref{fig_lattice}. Then the transformation $z^b_\l$ in \eqref{1-form_Z} is generated by conjugating $b_\l$ with $g^a_{\l^\star}$ whose $\l^\star$ is dual to $\l$ is the aforementioned manner (actually $\l^\star$ is right on top of $\l$, so we may therefore say $\l^\star=\l$); likewise the transformation $z^a_{\l^\star}$ is generated by conjugating $a_{\l^\star}$ with $g^b_{\l }$. The expressions of $g^a, g^b$ invite us to think of $\theta^a, \theta^b$ as the ``superconducting phases'' (with charge $n$) associated with the effectively $U(1)$ dynamical gauge fields $a, b$, respectively. Indeed, the expressions of $g^a, g^b$ are invariant under the ordinary ($0$-form) gauge transformations
\begin{align}
& b_\l \: \rightarrow \: b_\l + (\d\varphi^b)_\l, \ \ \ \ \ \theta^b_\v \: \rightarrow \: \theta^b_\v + n\varphi^b_\v \nonumber \\[.1cm]
& a_{\l^\star} \: \rightarrow \: a_{\l^\star} + (\d^\star\varphi^a)_{\l^\star}, \ \ \ \ \ \theta^a_{\v^\star} \: \rightarrow \: \theta^a_{\v^\star} + n\varphi^a_{\v^\star}, 
\label{0-form_U1}
\end{align}
where $\varphi^b_\v$, $\varphi^a_{\v^\star}$ are arbitrary $U(1)$ transformations on the vertices and dual vertices respectively; they are effectively $U(1)$ instead of $\mathbb{R}$ because any $2\pi\mathbb{Z}$ part of them can be completely absorbed into \eqref{1-form_Z}. The two transformations by $\varphi^b_\v$, $\varphi^a_{\v^\star}$ are, in turn, generated by the commutators with $f^a_{\p^\star=\v}$ and $f^b_{\p=\v^\star}$ respectively. In fact, it is easy to check that the generators $g^a, g^b, f^a, f^b$ for the gauge transformations \eqref{1-form_Z} and \eqref{0-form_U1} all commute with each other (so in particular their own expressions are invariant under \eqref{1-form_Z} and \eqref{0-form_U1}); if a state is a simultaneous eigenstate of all of $g^a, g^b, f^a, f^b$, we may view the state as respecting the respective Gauss's law constraints with the ``gauge charges'' given by the simultaneous eigenvalues.

\begin{figure}[t]
\includegraphics[width=.45\textwidth]{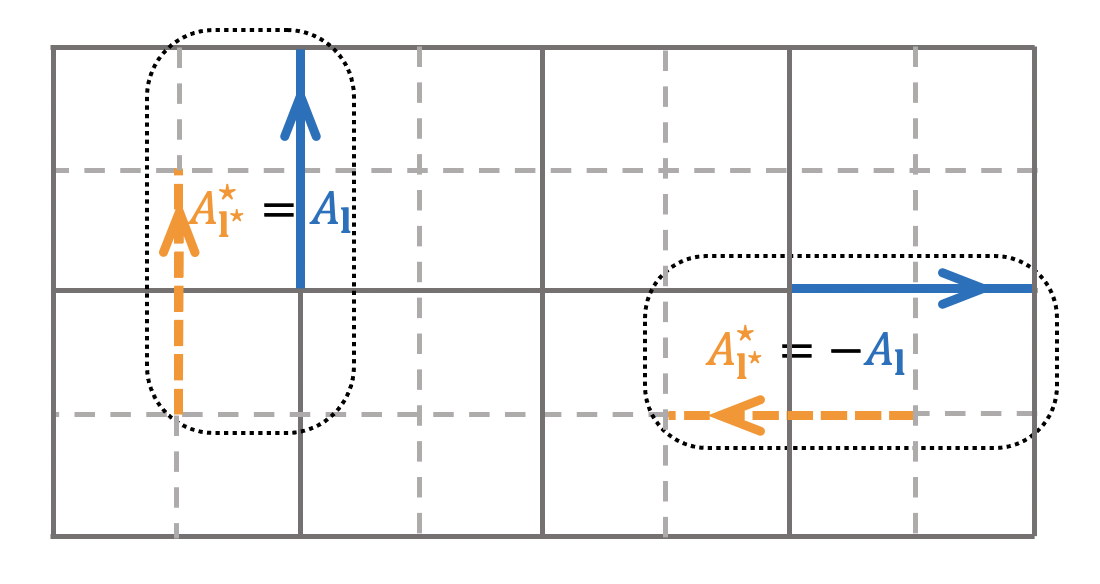}
\caption{A choice of identification of $A^\star_{\l^\star}$ with $A_\l$, where $\l$ is chosen to locate at $\hat{\mathbf{x}}/2+\hat{\mathbf{y}}/2$ away from $\l^\star$.}
\label{fig_lattice_A}
\end{figure}

Having introduced the local Hilbert space and operators, we now introduce the lattice Hamiltonian. Before we present our final Hamiltonian, it turns out to be helpful to first consider a simpler ``prototype'' Hamiltonian, and understand its properties and problems. Through the aforementioned effective Lagrangian and treatments that are detailed in Appendix \ref{app_L_to_H}, we are led to consider the ``prototype'' Hamiltonian
\begin{align}
\wt{H} \ =& \ \frac{U_b}{2}\sum_\l \left| 1 - e^{iqA_\l} g^b_\l \right|^2 + \frac{U_a}{2}\sum_{\l^\star} \left| 1 - e^{ipA^\star_{\l^\star}} g^a_{\l^\star} \right|^2 \nonumber \\[.2cm]
& + \ \frac{V_b}{2}\sum_\p \left(f^b_\p\right)^2 + \frac{V_a}{2} \sum_{\p^\star} \left(f^a_{\p^\star}\right)^2
\end{align}
where $A_\l$ is the electromagnetic $U(1)$ background field living on the link $\l$, while $A^\star_{\l^\star}$ on the dual link $\l^\star$ is identified with a nearby $A_\l$ pointing in the same direction, see Fig.~\ref{fig_lattice_A} for a choice of identification. The coupling of the electromagnetic field $A$ into the system can be intuitively understood as the following. If we think of $s^a_{\p^\star}$ as a boson number operator on the vertex $\v=\p^\star$, then $e^{\pm i\theta^b_\v}$ is the creation/annihilation operator of the boson, and hence $g^b_\l$ involves the hopping of such a boson across the link $\l$; if the boson carries electric charge $q$, its hopping will indeed couple to $A$ through the factor $e^{iqA_\l}$. Likewise for the term with $A^\star_{\l^\star}$. Therefore, the local electric charge operator on a vertex $\v$ is given by
\begin{align}
\rho_\v \ \equiv \ q \, s^a_{\p^\star=\v} + p \, s^b_{\p=\v-\hat{\mathbf{x}}/2-\hat{\mathbf{y}}/2}
\label{charge_density}
\end{align}
where the second term is due to the said identification between $A^\star_{\l^\star}$ and $A_\l$.

Let us examine the properties of $\wt{H}$. Since $g^a, g^b, f^a, f^b$ commute with each other, all of the four terms in $\wt{H}$ commute with each other and can be simultaneously diagonalized; as mentioned before, the simultaneous diagonalization may be viewed as imposing the Gauss's laws for their respective gauge transformations. But the four terms cannot be simultaneously minimized in general. To minimize the $U_b$ term, we need an eigenstate such that
\begin{align}
nb_\l = q A_\l + (\d \theta^b)_\l \ \mod 2\pi,
\label{U_min}
\end{align}
but then in the $V_b$ term
\begin{align}
f^b_\p = \frac{q}{n} (\d A)_\p \ \mod \frac{2\pi}{n}
\label{V_min}
\end{align}
which is non-zero in general, and hence the $V_b$ term is not minimized; in particular, for small magnetic field $\d A$, if the $U$ term is already minimized by the state, $f^b_\p = (q/n) (\d A)_\p$ is the choice that minimizes the $V$ term the best. Likewise for the $U_a$ term and the $V_a$ term, with $q\rightarrow p$ and $A\rightarrow A^\star$. Suppose $U_b/V_b, U_a/V_a \rightarrow \infty$ so that the $U_b, U_a$ terms are minimized first, and suppose the magnetic field $\d A$ is indeed small. Then the ground state may be viewed as considering all the eigenstates in the $b, \theta^b, s_b$ basis that first minimizes the $U_b$ term and then the $V_b$ term, and then taking a suitable linear superposition of all such states according to minimizing the $U_a$ term and then the $V_a$ term, in a spirit similar to solving for the toric code ground state(s) \cite{Kitaev:1997wr}; clearly one can also exchange the views between $b$ and $a$. 

Some connection to a $-2pq/n$ Hall conductivity can be readily noted. From the definitions of $f^b, f^a$ and $\rho$, we find the ground state expectation value of the local charge:
\begin{align}
\langle \rho_\v \rangle &= \ -\frac{q}{2\pi} \langle f^a - \d^\star a \rangle_{\p^\star} - \frac{p}{2\pi} \langle f^b - \d b \rangle_{\p} \nonumber \\[.2cm]
&= \ -\frac{pq}{2\pi n} \left( (\d A)_{\p'} + (\d A)_\p \right) \nonumber \\[.1cm] & \phantom{=} \ \ + \frac{q}{2\pi} \langle (d^\star a)_{\p^\star} \rangle + \frac{p}{2\pi} \langle (db)_\p \rangle
\end{align}
where $\p^\star = \v$, $\p=\v-\hat{\mathbf{x}}/2-\hat{\mathbf{y}}/2$, $\p'=\v+\hat{\mathbf{x}}/2+\hat{\mathbf{y}}/2$. Since $(\d A)_\p$ and $(\d A)_{p'}$ are nearby magnetic fields, this result already has the flavor of $\langle \delta \rho \rangle = \sigma_H \d A$ with the Hall conductivity $2\pi \sigma_H = -2pq/n$, if we can show that $\langle (d^\star a)_{\p^\star} \rangle$ and $\langle (db)_\p \rangle$ at ground state are somehow independent of the external magnetic field.

At this point we are ready to see the problem with our ``prototype'' Hamiltonian $\wt{H}$. The problem manifests itself as two-folded:
\begin{enumerate}
\item Since $g^a, g^b, f^a, f^b$ commute with each other and their eigenvalues are all continuous, the Hamiltonian is gapless, rather than a gapped one that we want -- in fact, the Hamiltonian is locally gapless, not just becoming gapless only after taking the limit of large system size. Related to this, the ground state and hence the response to the magnetic field are sensitive to the ratios $U_b/V_b$ and $U_a/V_a$ (which were taken towards infinite in the above), rather than being robustly fractionally quantized. 
\item
To compute the Hall conductivity, we need to show that the ground state local expectations values $\langle (d^\star a)_{\p^\star} \rangle$ and $\langle (db)_\p \rangle$ are somehow independent of the external magnetic field. But with a little extra effort it is easy to see these expectation values are ambiguous, somewhat like asking for the expectation $\langle x \rangle$ for a Bloch wave. The ambiguity arises from the infinite ranges of the local values of $a, b$ involved in the superposition.
\end{enumerate}
These two issues actually represent the same problem -- since $a, b$ are canonical conjugates, the local gaplessness of $a$ is related to the local unboundedness of $b$, and vice versa.

Our task is therefore to resolve this problem of $\wt{H}$, making a modified Hamiltonian gapped and its Hall conductivity unambiguous. This is the point at which the Kapustin-Fidkowski no-go theorem \cite{kapustin2020local} makes its manifestation. When the Hall conductivity vanishes, i.e. when $p=0$ or $q=0$, there is a known resolution that maintains the local commutativity of the terms in the Hamiltonian, making it exactly solvable; the model was introduced in \cite{Levin:2011hq} (as part of a larger Hamiltonian), and revisited in our present framework in \cite{Chen:2019mjw}, see Appendix \ref{app_L_to_H}. Roughly speaking, the resolution is to note that when $q=0$ (the case when $p=0$ is analogous), the condition \eqref{U_min} is independent of the external field, and therefore, rather than viewing it as an energetic condition, we may instead, from the very beginning, let the full local Hilbert space on the link to take a finite set of discrete values $\wt{b}_\l=\{ 0, 1, \cdots, n-1 \}$ (which may be viewed as a reminiscence of $(n b-\d \theta^b)_\l /2\pi \mod n$ under the condition \eqref{U_min}); now that the local Hilbert space on a link is discrete and bounded, one can construct gapped local commuting terms that essentially play the roles of the $V_b, U_a, V_b$ terms in $\wt{H}$, and the commutativity leads to solvability. If both $p, q=0$, the model can be further reduced to the $\mathbb{Z}_n$ generalization of the toric code \cite{Kitaev:1997wr}. Such a model becomes unavailable when the Hall conductivity $-2pq/n\neq 0$, i.e. when $p, q \neq 0$, since the condition \eqref{U_min} and its $a_{\l^\star}$ analogue both depend on the external field and can no longer be realized by any fixed discrete local Hilbert space to begin with. Thus, for general values of $p, q$, we will resort to our resolution below, which makes the Hamiltonian not exactly solvable, but fortunately perturbatively solvable at its lower energy sectors.

Our resolution is to add simple, non-commuting terms $\epsilon_a a_{\l^\star}^2/2$ and $\epsilon_b b_{\l}^2/2$ to the ``prototype'' Hamiltonian $\wt{H}$, to open the local gaps and also to softly bound the ranges of the local Hilbert spaces. The Hamiltonian reads
\begin{align}
H =& \ \sum_{\l = \l^\star} \left[ \ \frac{\epsilon_a}{2} a_{\l^\star}^2 + \frac{\epsilon_b}{2} b_\l^2 \right. \nonumber \\[.1cm] & \ \ \ \ \ \left. + \ U_b\left( 1 - \cos(\d\theta^b - nb + qA)_\l \right) \right. \nonumber \\[.1cm] & \ \ \ \ \ \left. + \ U_a\left( 1 - \cos(\d^\star\theta^a - na + pA^\star)_{\l^\star} \right) \phantom{\frac{1}{1}} \right] \nonumber\\[.2cm]
& \ + \sum_{\p=\v^\star} V_b \left( \d b - 2\pi s^b \right)_\p^2 \nonumber \\[.1cm] & \ +  \sum_{\v=\p^\star} V_a \left( \d^\star a - 2\pi s^a \right)_{\p^\star}^2 \ .
\end{align}
Since the $\epsilon_a, \epsilon_b$ terms do not commute with the some of the remaining terms (nor with each other), a gap is opened up; the fluctuations of $a, b$ on each link are also softly bounded by these terms. As the $\epsilon_a, \epsilon_b$ terms violate the gauge invariances \eqref{1-form_Z}, \eqref{0-form_U1}, the gauge field picture of the $a, b$ variables is no longer exact, but emergent at best. Indeed, in the remaining sections of this paper, we will show such a gauge field picture emerges at the low energy sector on each individual link when $\epsilon_a, \epsilon_b$ are small compared to $U_a, U_b$, and we are able to solve the low energy sectors of the full Hamiltonian, and show the system exhibits the desired Hall conductivity $-2pq/n$.

Before we move on towards the solution, we would like to comment on the relation between our work and the relevant literature. We first remark that the Hamiltonian $H$ has appeared before as certain cases in \cite{geraedts2013exact,geraedts2017lattice}, which numerically studied lattice models with Hall conductivity. The proper relation of the lattice models to Chern-Simons theory and the subsequent path towards analytically solving the models, however, were not addressed in \cite{geraedts2013exact,geraedts2017lattice}. In Appendix \ref{app_relation} we provide a detailed explanation of how to properly identify the nature of the topological orders for all of the models in \cite{geraedts2013exact,geraedts2017lattice}, so to connect those models to Chern-Simons theory and to our work. The novelty of our work is that we motivated the lattice Hamiltonian $H$ starting from a class of well-established Chern-Simons theories, from which the solvability follows as a natural consequence. Our approach is systematic and can be straightforwardly applied to the more general twisted bosonic and fermionic topological orders \cite{Chen:2019mjw}. We will elaborate on those more general cases in upcoming works; the present work serves to demonstrate our approach through the most basic examples.

In \cite{geraedts2017lattice} the relation between some of the models and the $\mathbb{Z}_n$ toric code was discussed in the absence of coupling to the electromagnetic background (see Appendix \ref{app_relation}). As we will see in the below, interestingly, when we solve the Hamiltonian in the presence of electromagnetic coupling, crucial novel features appear: the low energy subspace emerges to be an effective $\mathbb{Z}_n$-gauge-like theory, but with the $\mathbb{Z}_n$ in some ways ``shifted by'' and ``projective under'' the electromagnetic background; such modification is essential for a non-trivial Hall conductivity. Such background dependent emergent field is beyond the celebrated paradigm of exactly solvable lattice models established since the seminal works \cite{Kitaev:1997wr, Levin:2004mi, Chen:2011pg}, and may shed light on the studies of the more general symmetry enriched topological orders that are beyond this celebrated paradigm.

More recently \cite{DeMarco:2021erp} investigated exactly solvable local commuting projector Hamiltonians with integer values of Hall conductivity. There are two differences with our present work. First, \cite{DeMarco:2021erp} focused on integer Hall conductivity, while the Hall conductivity in our present work takes fractional values in general. Second, the spirit of their lattice model bears a major difference with ours. The model in \cite{DeMarco:2021erp} involved $U(1)$ valued local Hilbert space, and for their Hamiltonians to be made of local commuting projectors, the matrix elements must not be continuous functions of both the $U(1)$ dynamical variables and the electromagnetic $U(1)$ background gauge field. On the other hand, our Hamiltonians' matrix elements are regular functions of the continuous dynamical variables; the electromagnetic $U(1)$ background gauge field also couples to the system in the usual form. Our models hold on to these important physical requirements \cite{kapustin2020local} at the cost of exact solvability. It is interesting to note, however, that the discontinuities in the models of \cite{DeMarco:2021erp} may be interpreted as a reminiscence of the emergent $1$-form $\mathbb{Z}$ gauge invariance \eqref{1-form_Z}; see the discussion below \eqref{bbaseexpansion} for the origin of the emergent ``discontinuity'' in our continuous model. Since it is well-known that discontinuities appear when the group cohomology machinery \cite{Chen:2011pg} is applied to continuous global symmetry groups (electromagnetic $U(1)$ here), our resolution to the discontinuity problem is another reflection that our construction is reaching beyond the currently established paradigm.

\section{Solving the Hamiltonian}

\begin{figure}[t]
\includegraphics[width=.45\textwidth]{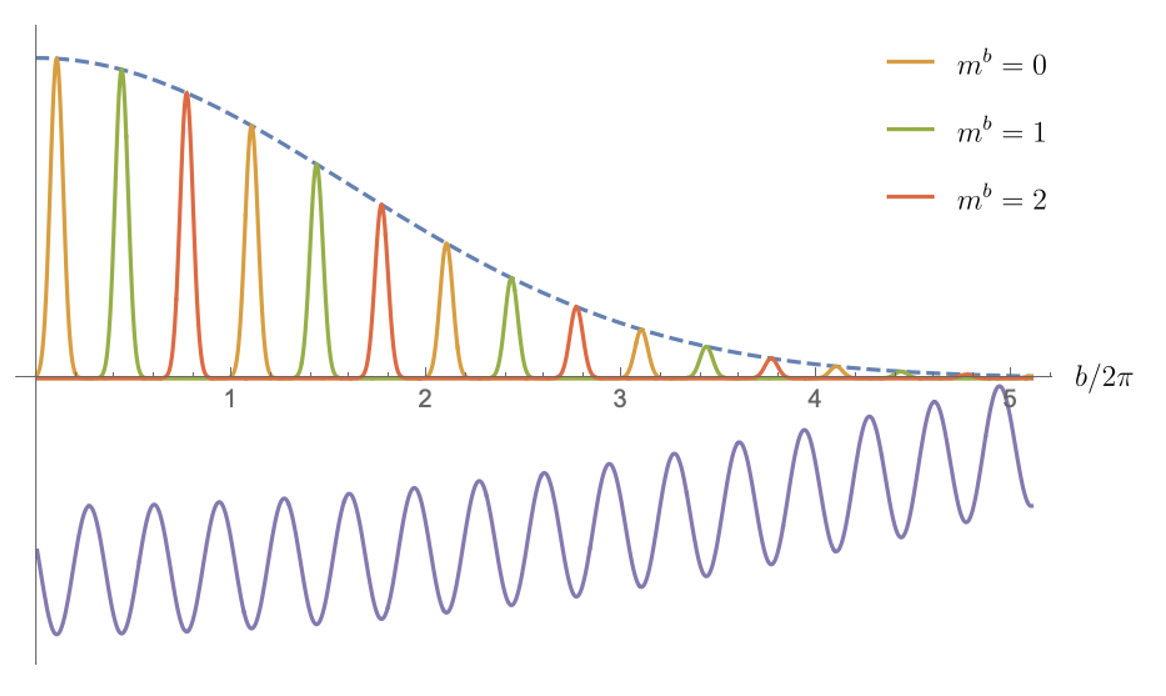}
\caption{The trial ground state wavefunctions $|{\Psi}^{(0,0)}_{m^b}(b)|$ for the local Hamiltonian on each link, in the $b$ basis. The ``potential'' for the $b$ variable, $(\epsilon/2)b^2 - U \cos n(b-b_0)$, is illustrated below with the purple line. The parameters used are $n=3$, $\epsilon_b = \epsilon_a = \epsilon$, $U_a = U_b = U$, $\epsilon/U=1/4^4$, and $b_0=0.2\pi$.  In the $b$ basis, the wide envelope represented by the dashed line is a Gaussian centered at $0$, with width $2\pi/nW_a$ where $W_a\sim(\epsilon_b/U_a)^{1/4}$; each narrow peak is a Gaussian centered at $\bar{b}_j=(2\pi/n)j+\bar{b}_0$, with width $W_b\sim(\epsilon_a/U_b)^{1/4}$. The errors of these trial wavefunctions to the true ones are $\mathcal{O}(W^2)$.}
\label{fig_gsonlink}
\end{figure}

Now we proceed to solve the Hamiltonian $H$ at its low energy sector and find the Hall conductivity. The sketch is the following. 
\begin{enumerate}
\item
First we solve the link terms in a solvable limit, i.e. perturbatively find the low energy subspace determined by the $\epsilon_a, \epsilon_b, U_a, U_b$ terms on each individual link, which turn out to be an emergent $\mathbb{Z}_n$ subspace but with operator values shifted by the electromagnetic field. The approximate low energy wavefunctions are illustrated in Fig.~\ref{fig_gsonlink}. 
\item
Then, with this emergent $\mathbb{Z}_n$ low energy subspace on each link, we can solve the $V_a, V_b$ terms in a manner similar to solving the toric code \cite{Kitaev:1997wr}, and find fractionally charge anyons and the fractional Hall conductivity. 
\item
Finally, since we have used perturbation theory, we need to confirm that the Hall conductivity is precisely the designated fraction $-2pq/n$, rather than some nearby number blurred by the perturbative error. This is done by narrowing down our perturbative error without closing the gap, and implementing the usual derivation for the robustness of fractional Hall conductivity in gapped systems \cite{niu1985quantized}.
\end{enumerate}

In this and the following section we carry out these steps. We remark that our solving procedure resembles what one does to study a real solid state or cold atom system: one first identifies the local low energy subspace -- the relevant electronic orbits on an atom, or the relevant atomic orbits in a potential trap -- out of an infinite dimensional local full Hilbert space; then one projects the many-body interactions onto the low energy subspace to study the many-body problem. 

\subsection{Local Low Energy Subspace}
\label{sect_LLES}

In this subsection, we solve the local low energy subspace on each link in a solvable limit. The local Hamiltonian can be expressed as (the link index is omitted in this section):
\begin{align} \label{localH}
H_{\text{link}} &= \frac{1}{2} \left(\epsilon_b b^2 +\epsilon_a a^2 \right) \nonumber \\[.2cm] & \ \ \ \ - U_b \cos[n(b-\bar{b}_0)] - U_a \cos[n(a - \bar{a}_0)]
\end{align}
where $\bar{a}_0$ and $\bar{b}_0$ include terms that commute with the $b$, $a$ operators. Our task is to perturbatively find the low energy subspace of $H_{\text{link}}$, under the assumption
\begin{align}
\epsilon \ll U,
\end{align}
where $\epsilon$ is the scale of $\epsilon_a$ and $\epsilon_b$, and $U$ the scale of $U_a$ and $U_b$. We will find trial wavefunctions for $n$ nearly degenerate low energy states that are well separated for other states, forming an emergent $\mathbb{Z}_n$ low energy subspace; the actual low energy wavefunctions differ from our trial ones by $\mathcal{O}(\sqrt{\epsilon/U})$ controlled errors, and their energy split is $\mathcal{O}(\sqrt{\epsilon/U})$ smaller than their gap to the higher energy states. The full details of the calculation and error control in this section can be found in Appendix \ref{app_LLES}.

To motivate our trial wavefunctions, note that the leading non-commuting pairs of terms are the $\epsilon_a$ and $U_b$ pair, as well as the $\epsilon_b$ and $U_a$ pair. Let us first consider the $\epsilon_a$ and $U_b$ terms
\begin{align}
    H_{\text{link},b} = \frac{1}{2}\epsilon_a a^2  - U_b \cos[n (b-\bar{b}_0)]
\end{align}
which may be viewed as a particle in a one-dimensional sinusoidal potential. Alternatively, if we are to view $b$ as some gauge field as motivated in the previous section, then the $U_b$ term is the Higgs potential and the $\epsilon_a$ term is the kinetic energy that opens the Higgs gap near the potential mimima. As long as the potential is deep enough, the neighborhood around each minimum, located at $\bar{b}_j = (2\pi/n)j + \bar{b}_0$ with $j\in \mathbb{Z}$, can be effectively expanded to the quadratic order. Therefore, the low energy eigenfunctions are approximately those of infinitely many harmonic oscillators, one at each potential minimum $\bar{b}_j$. The eigenstates $|\phi^{(N_b)}_{j}\rangle$ are those of the harmonic oscillators at the potential minima $\bar{b}_j$ and with energy level $N_b$. The excitation energy of which is
\begin{align}
    \omega_b = 2\pi \sqrt{U_b\epsilon_a}.
\end{align}
In particular the low energy $|\phi^{(N_b)}_{j}\rangle$ are localized Gaussians of width
\begin{align}
    W_b = \frac{\sqrt{2\pi}}{n} \left(\frac{\epsilon_a}{U_b}\right)^{\frac{1}{4}}
\end{align}
centered at $\bar{b}_j$. As long as $N_b$ is not too large, the tunneling overlap between different $j$s is exponentially small in $-1/W_b^2$, hence the $N_b$ states, at least for small $N_b$ can be treated as a nearly orthonormal and nearly degenerate under $H_{\text{link},b}$, analogous to a ``tight binding'' electron in a sinusoidal potential. We need to consider $N_b=0$ for low energy subspace, and small $N_b>0$ for error control. The higher powers in the expansion of the $\cos$ minima mixes $N_b=0$ with small $N_b$ and modifies the wavefunctions by a controlled error $\mathcal{O}(W_b^2)$.

Now we have a new set of basis states describing the low energy physics of the system, spanned by infinitely many highly localized orbits $|\phi_j^{(N_b)}\rangle$. Our next step would be to solve the remaining terms,
\begin{align}
    H_{\text{link},a} = \frac{1}{2}\epsilon_b  b^2 - U_a \cos[n (a-\bar{a}_0)]
\end{align}
in this basis. While the $U_a$ term, the ``hopping term of the tight binding model'', remains exact after the projection into this basis (as this basis spans an invariant subspace under the $U_a$ term), the ``potential well'' $\epsilon_b$ term does not. We approximate 
\begin{align} \label{approxbbar}
    b^2 |\phi_j^{(N_b)} \rangle  \approx \bar{b}^2 |\phi_j^{(N_b)} \rangle
\end{align}
which we must justify later (see the end of this subsection). Here $\bar{b}$ is the operator identifying the center coordinates $(2\pi/n)j + \bar{b}_0$ of the orbits, which is well-defined at least for $N_b$ not too large. This means that at low energies we no longer need to consider the continuous $b$ variable but can concentrate on a discrete $\bar{b}$ space for each level $N_b$. The basis of the conjugate variable of $\bar{b}$ operator, $\underline{a}$ (not to be confused with $\bar{a}$), is then confined onto the ``first Brillioun zone'' in the reciprocal space of the $\bar{b}_j$ lattice, i.e. $\underline{a} \in [0,2\pi)$. Since the roles of $\underline{a}$ and $\bar{b}$ are conjugate, we can switch the perspective and now interpret $\underline{a}$ as a coordinate on a ring. Thus we can regard $-U_a \cos[n (a-\bar{a}_0)]=-U_a \cos[n (\underline{a}-\bar{a}_0)]$ term as a sinusoidal potential, subjected to the periodic boundary condition identifying $\underline{a}=0$ and $2\pi$. This potential has $n$ minima located at $\bar{a}_{m^a} = (2\pi/n)m^a +\bar{a}_0$, $m^a\in \mathbb{Z}_n$. We can solve the Hamiltonian around each minimum, again treating the potential in the neighborhoods of the minima as quadratic. The corresponding excitation energy is
\begin{align}
    \omega_a = 2\pi \sqrt{\epsilon_b U_a}
\end{align}
and the low energy states, $\tilde{\Psi}^{(N_b,N_a)}_{m^a}$ can be solved. They are $n$ fold degenerate for each level, and again localized in $\underline{a}$ basis by Gaussian width
\begin{align}
    W_a = \frac{\sqrt{2\pi}}{n} \left(\frac{\epsilon_b}{U_a} \right)^{\frac{1}{4}}.
\end{align}
Similar to the discussion in the previous step, the errors are $\mathcal{O}(W_a^2)$.

Piecing up the above, we conclude that the link has an emergent $\mathbb{Z}_n$-like low energy subspace, with trial wavefunctions
\begin{widetext}
\begin{align}\label{gslinktransformedina}
      |\tilde{\Psi}^{(0,0)}_{m^a}\rangle &=  \sqrt{\frac{W_b}{n\pi W_a} }  \int da \ \mathrm{e}^{-\i\frac{n}{2\pi} \bar{b}_0 (a-\bar{a}_{m^a})}\mathrm{e}^{-W_b^2(n/2\pi)^2a^2/2}    \left(  \sum_{z^a\in \mathbb{Z}}  \mathrm{e}^{-[a-\bar{a}_{m^a+nz^a}]^2/2W_a^2} \right) |a \rangle 
\end{align}
or their linear combinations
\begin{align}\label{gslink}
|{\Psi}^{(0,0)}_{m^b}\rangle &\equiv \frac{1}{\sqrt{n}}\sum_{m^a} \mathrm{e}^{-\i\frac{n}{2\pi}\bar{a}_{m^a}\bar{b}_{m^b}} |\Psi^{(0,0)}_{m^a}\rangle \nonumber \\[.2cm]
    &=   \sqrt{\frac{W_a}{n\pi W_b} } \int db \left( \sum_{z^b\in \mathbb{Z}}  \mathrm{e}^{\i n \bar{a}_0 z^b} \mathrm{e}^{-(n/2\pi)^2W_a^2\bar{b}_{m^b+nz^b}^2 /2}   \mathrm{e}^{-[b-\bar{b}_{m^b+nz^b}]^2/2W_b^2} \right)|b \rangle \ .
\end{align}
\end{widetext}
Here $m^a$ and $m^b$ take values in $\{0, 1, \cdots, n-1\}$; if we shift either by $n$, the corresponding wavefunction will return to itself but with an overall phase that depends on $\bar{b}_0$ or $\bar{a}_0$, an important feature different from an actual $\mathbb{Z}_n$, as we will see later. An illustration of ${\Psi}^{(0,0)}_{m^b}$ in $b$ basis can be found in Fig.~\ref{fig_gsonlink}; it is clear that the solutions to $H_{\text{link},b}$ give the narrow peaks, with width $W_b$, while the solution to $H_{\text{link},a}$ determines the broad envelope, with width $2\pi/ nW_a$.

One may notice the two expressions above are slightly asymmetric between $a$ and $b$. This is because we treated $H_{\text{link}, b}$ first. There is no contradiction to the apparent symmetry between $a$ and $b$ in the original problem (if we have set $\epsilon_a=\epsilon_b=\epsilon$ and $U_a=U_b=U$), because these are trial wavefunctions that have $\mathcal{O}(\sqrt{\epsilon/U})$ errors with the actual low energy states anyways, and the said asymmetry is indeed of this order.

At this point we shall return and justify the approximation in \eqref{approxbbar}. At first sight it seems this cannot be justified because no matter how small $\epsilon_b/U_b$ is, the $b^2$ potential will eventually be large enough to overcome the $\cos$ potential which determines the $\bar{b}$ minima; in other words, it is not obviously controlled to find the low energy states of $H_{\text{link}, b}$ first and then project $H_{\text{link}, a}$ into them. Our justification, in intuitive terms, is to note that the trial wavefunctions $|{\Psi}^{(N_a, N_b)}_{m^b}\rangle$ (with small values of $N_a, N_b$ of interest) already have small amplitudes when $b^2$ is large enough to overcome $\epsilon_b/U_b$. In particular, the typical spread of $b$ in these wavefunctions is determined by the broad envelope of width $\sim1/W_a \sim (\epsilon/U)^{-1/4}$, therefore $\epsilon_b b^2 \sim \epsilon \sqrt{U/\epsilon}$ is still smaller than $U_b$ by a factor of $\mathcal{O}(\sqrt{\epsilon/U})$. A more rigorous error control is given in Appendix \ref{app_LLES}. This error may introduce splitting among the $n$ actual low energy states, by an amount $\mathcal{O}(\epsilon)$ which is $\mathcal{O}(\sqrt{\epsilon/U})$ smaller than the gap $\mathcal{O}(\sqrt{\epsilon U})$ with the higher energy states.

\subsection{Many-Body Ground States and Excitations}
\label{sect_solution}

Now that we have found the link low energy subspace to be an effective $\mathbb{Z}_n$-like field (but bears important difference with a literal $\mathbb{Z}_n$ field, see discussion below \eqref{bbaseexpansion}), one can envision that the remaining solution to the many-body problem is similar to that of a $\mathbb{Z}_n$ toric code, with suitable modifications by the electromagnetic background. We will see this is indeed the case. To control the error, the assumptions to be made are
\begin{align}
\epsilon \ll V \ll \sqrt{\epsilon U}
\end{align}
where $V$ is the scale of $V_a, V_b$. The second separation of scales is to ensure the applicability of the usual many-body perturbation theory, where we can neglect the higher energy states in each local Hilbert space when considering the interaction couplings between local systems; this is justified in Appendix \ref{app_V_terms_control}. The first separation of scales is to ensure that the energy split between the states within each local low energy subspace can be treated as a local perturbation on top of the toric code like physics, so that the robustness of the toric code like physics \cite{Kitaev:1997wr} is applicable. 

We first note that for each link low energy trial state, we shall make the substitutions
\begin{align}
    b_{0,\l} \rightarrow (\d\theta^b +q A)_\l/n &, \ \ a_{0,\l^\star} \rightarrow (\d^\star\theta^a +p A^\star)_{\l^\star}/n \nonumber
\end{align}
so that the link state $|{\Psi}^{(0,0)}_{m^b, \l}\rangle$ depends on the neighboring vertex states $|\theta^b_\v\rangle$ and plaquette states $|\theta^a_{\v^\star}\rangle$, and also on the electromagnetic background. So any given sets of $\{m^b_\l\}$ , $\{\theta^a_{\v^\star}\}$, $\{ \theta^b_\v \}$ specify a basis state for the many-body low energy basis,
\begin{align}\label{fourierbasis}
    |\{m^b_\l\} , \{ \theta^a_{\v^\star}\}, \{ \theta^b_\v \}\rangle \equiv  \bigotimes_{\v^\star, \v, \l}  |\theta^a_{\v^\star} \rangle | \theta^b_\v \rangle |{\Psi}^{(0,0)}_{m^b_\l, \l}\rangle
\end{align}
The Fourier transforms of $\{ \theta^a_{\v^\star}\}$ can also serve as a well defined basis:
\begin{align}
&|\{m^b_\l\} , \{s^b_\p\}, \{ \theta^b_\v \} \rangle \nonumber\\
\equiv& 
\int_{\{\theta^a_{\v^\star}\}} \ \mathrm{e}^{\i\sum_{\v^\star} s^b_\p\cdot \theta^a_{\v^\star}} |\{m^b_\l\} , \{ \theta^a_{\v^\star}\}, \{ \theta^b_\v \}\rangle  \label{bbasedefinition}
\end{align} 
The basis vectors can be expanded in the original basis (for simplicity, we omit the possible normalization factors in this subsection)
\begin{widetext}
\begin{align}
|\{m^b_\l\} , \{s^b_\p\}, \{ \theta^b_\v \} \rangle =& \sum_{\{z^b_\l\}} \int_{\{b_\l\}} \left(\prod_\l
\ \mathrm{e}^{\i z^b_\l \cdot (p A^\star)_{\l^\star}} \mathrm{e}^{-\bar{b}_{m^b_\l,z^b_\l}^2(n/2\pi)^2W_a^2/2}  \ \mathrm{e}^{-(b_\l-\bar{b}_{m^b_\l,z^b_\l})^2/2W_b^2} \right) \bigotimes_{\l,\p,\v} |b_{\l}\rangle  |(s^b + \d z^b)_\p \rangle | \theta^b_\v \rangle  \label{bbaseexpansion}
\end{align} 
\end{widetext}
where the minima positions
\begin{align}
\bar{b}_{m^b_\l,z^b_\l} \equiv \frac{2\pi}{n} \left( m^b+nz^b+\frac{\d\theta^b+qA}{2\pi} \right)_\l.
\end{align} 
We make a few remarks about this set of many-body low energy basis states. 
\begin{itemize}
\item The local minima in the $b$ basis, specified by $\bar{b}_{m^b_\l,z^b_\l}$, may be viewed as an emergent realization of the Gauss's constraint $g^b_\l = e^{-iqA_\l}$ (see also \eqref{U_min}). The positions of the minima are shifted by the background electromagnetic field $A_\l$.
\item For definiteness of the labels, we need to fix the emergent $\mathbb{Z}_n$ label $m^b_\l$ to take values in the range, say, $\{0,1,\dots, n-1\}$. If an $m^b_\l$ shifts by $n$, the effect can be absorbed into a shift of the nearby $s^b_\p$ through a shift of the summation variable $z^b_\l$; moreover, an overall phase that depends on $A^\star_{\l^\star}$ is generated. In other words, for $\mathbb{Z}_n=\mathbb{Z}/n\mathbb{Z}$, the mod out of $n\mathbb{Z}$ is now ``projective'' under the electromagnetic field $A^\star_{\l^\star}$. It is easy to see that the summation over $z^b_\l$, entangling the link Hilbert space and the nearby plaquette Hilbert spaces, is an emergent realization of the Gauss's constraint $g^a_{\l^\star}=e^{-ipA^\star_{\l^\star}}$ which generates the first line of the invariance \eqref{1-form_Z}. 
\item While any final physical result must not depend on any $2\pi$ shift of $\theta^b_\v$ or $A_\l$ because the original Hamiltonian does not, the labeling of the basis states does. If such a shift is made, through the expression of $\bar{b}_{m^b_\l,z^b_\l}$ we can see $m^b_\l$ must be relabeled to represent the same physical state; furthermore, if $m^b_\l$ is relabeled out of the $\{0,1,\dots, n-1\}$ range, to relabel it back in, the nearby $s^b_\p$ must also be relabeled, and an overall phase that depends on (``projective under'') $A^\star_{\l^\star}$ is generated. One may avoid this ambiguity by fixing $\theta^b_\v$ and $A_\l$ to belong to $(-\pi , \pi]$, but we do not have to, as long as the said relabeling is understood.
\end{itemize}
All these are crucial features of our emergent $\mathbb{Z}_n$-like theory, compared to a usual lattice $\mathbb{Z}_n$ theory that cannot couple to electromagnetic background. In particular, the last point above is closely related to the \emph{effective} indistinguishability \eqref{eff_indistinguishability}, which is in turn crucial to the embodiment of the Kapustin-Fidkowski no-go theorem, see Appendix \ref{app_L_to_H}. More broadly, the fact that the emergent $\mathbb{Z}_n$-like field is ``shifted by'' and ``projective under'' the electromagnetic background is a novel feature that cannot be achieved in the standard paradigm of constructing exactly solvable lattice models \cite{Kitaev:1997wr,Levin:2004mi,Chen:2011pg} which would have started with a literal $\mathbb{Z}_n$ field; in particular, the relabeling of $m^b_\l, s^b_\p$ needed when $A_\l$ gradually increases from $0$ to $2\pi$ may be interpreted as the ``discontinuity'' \cite{DeMarco:2021erp} in the effective description -- and here we see such ``discontinuity'' emerges from a lattice model that is in fact continuous in all variables. This may shed light on the study of more general symmetry enriched topological orders.

It is useful to also introduce the dual basis states $|\{{m^a_{\l^\star}}\} , \{\theta^a_{\v^\star}\}, \{ s^a_{\p^\star} \} \rangle$ in a symmetric manner to \eqref{bbaseexpansion}, with minima
\begin{align}
\bar{a}_{m^a_{\l^\star},z^a_{\l^\star}} \equiv \frac{2\pi}{n} \left( m^a+  nz^a +  \frac{\d^\star\theta^a+pA^\star}{2\pi} \right)_{\l^\star} .
\end{align}
(As commented in the paragraph after \eqref{gslink}, within our controlled error we can safely neglect the slight asymmetry between the $\langle a |\tilde{\Psi}^{(0,0)}_{m^a}\rangle$ and the $\langle b |\Psi^{(0,0)}_{m^b}\rangle$ expressions.) The formal remarks are analogous to the above and will not be repeated.

The many-body part of the Hamiltonian is $H_b+H_a$, where $H_b \equiv V_b \sum_\p ( \frac{\d b}{n} - 2\pi s^b)^2_\p$ and $H_a \equiv V_a \sum_\v ( \frac{\d^\star a}{n} - 2\pi s^a)^2_{\p^\star}$. In Appendix \ref{app_V_terms_control} we justify that it suffices to focus on their projections into the local low energy trial subspace we found above. As in the usual toric code these two terms commute and can be simultaneously diagonalized, even so after the projection into the local low energy trial subspace, up to errors exponentially small in $1/W^2$. Moreover, up to the same error, it is straightforward to show that $H_b$ and $H_a$ are respectively diagonal in the two sets of trial basis states introduced above, with eigenvalues
\begin{align}
& V_b \sum_\p \left( \frac{2\pi \d m^b+q\d A}{n} - 2\pi s^b \right)^2_\p \label{eigenvaluesb}, \\[.2cm]
& V_a \sum_{\p^\star} \left( \frac{2\pi \mathrm{d^\star}m^a+p\d^\star A^\star}{n} - 2\pi s^a \right)^2_{\p^\star} \label{eigenvaluesa}
\end{align}
respectively. Therefore we can denote the local excitation numbers by
\begin{align} 
& v^b_\p \equiv -\left(\d m^b-ns^b + [\frac{q\d A}{2\pi}] \right)_\p, \label{excitation_b} \\[.2cm]
& v^a_{\p^\star} \equiv -\left(\d^\star m^a-ns^a + [\frac{p\d^\star A^\star}{2\pi}] \right)_{\p^\star}, \label{excitation_a}
\end{align}
where $[x]$ denotes the nearest integer to a real number $x$. $H_b$ and $H_a$ are minimized when $v^b$ and $v^a$ vanish respectively.

To see $v^b_\p$ describes anyon excitations, we use the Wilson loop operator
\begin{align}
L^b_\ell &\equiv \ \exp\left[ i \sum_{\l \in \ell} b_\l \right] \rightarrow \ \exp \left[ i\frac{2\pi}{n} \sum_{\l \in \ell} \left(m^b + q \frac{A}{2\pi} \right)_\l \right]
\label{Lb_loop}
\end{align}
with $\ell$ a lattice loop encircling $\p$; the expectation would be a phase $e^{-\i 2\pi v^b_\p/n}$ in the absence of background field. Moreover, it is easy to see that if $\ell$ is a line with two open ends, the Wilson line $L^b_\ell$ will create a $\pm 1$ pair of $v^a_{\p^\star}$ excitations at its ends. This is why $-v^b$ and $v^a$ are usually called the ``flux'' and ``charge'' anyons in the $\mathbb{Z}_n$ toric code, named in the perspective of $b$ being a gauge field. The $-2\pi/n$ phase in the expectation value is the mutual braiding statistics between the charge and the flux anyons. Similarly, 
\begin{align}
L^a_{\ell^\star} &\equiv \ \exp\left[ i \sum_{\l^\star \in \ell^\star} a_{\l^\star} \right]  \rightarrow \ \exp \left[ i\frac{2\pi}{n} \sum_{\l^\star \in \ell^\star} \left(m^a + p \frac{A^\star}{2\pi} \right)_{\l^\star} \right]
\label{La_loop}
\end{align}
detects $v^a_{\p^\star}$ if $\ell^\star$ is a dual lattice loop encircling $\p^\star$, and creates a $\pm 1$ pair of $v^b_\p$ if $\ell^\star$ has open ends. We will leave the discussions of the electromagnetic responses to the next section.

Consider the condition \eqref{excitation_b}. Different $m^b, s^b$ configurations may have the same values of $(\d m^b-ns^b)_\p$. Locally, such equivalent configurations is generated by the $\mathbb{Z}/n\mathbb{Z}$ reminiscence of the $0$-form $\mathbb{R}/2\pi\mathbb{Z}$ gauge invariance, first line of \eqref{0-form_U1} rescaled by $n/2\pi$. The mod out of the $n\mathbb{Z}$ is embodied such that, if a $0$-form gauge transformation brings some $m^b_\l$ out of the specified range $\{0,1,\dots, n-1\}$, we must bring it back into the range by the $1$-form $\mathbb{Z}$ gauge transformation mentioned before, which thereby changes the nearby $s^b_\p$ configurations \cite{Levin:2011hq}. The explicit form of such local transformation is given in \eqref{app:Tdefinition} in Appendix \ref{app_solution}.

Globally, there are also $m^b, s^b$ configurations that give the same values of \eqref{excitation_b} everywhere, but nonetheless belong to different topological classes $C$. Consider the square lattice with periodic boundary conditions, forming a torus. Each topological class $C$ is characterized by the expectation values of two Wilson loop operators, $L^b_{\ell_x}$ and $L^b_{\ell_y}$, where $\ell_i$ is a non-contractible loop that runs around the $i$ direction. The particular path that $\ell_i$ runs through does not matter if $A_\l=0$ and $v^b_\p=0$ everywhere, but otherwise we need to fix the particular path to make comparisons. Both $L^b_{\ell_x}$ and $L^b_{\ell_y}$ can take $n$ different expectation values differing from each other by a $e^{i2\pi/n}$ phase. Therefore, for fixed $A$ and $v^b$, the total number of topological classes $C$ on a torus is $n^2$ \cite{Kitaev:1997wr}.

Similar discussions apply if we consider the condition \eqref{excitation_a}. The local transformation is given in \eqref{app:Tstardefinition}. The topological classes, characterized by $L^a_{\ell^\star_x}$ and $L^a_{\ell^\star_y}$, are labeled by $C^\star$.

So, far, we have considered $H_a$ and $H_b$ separately under the two sets of bases $\left\{ |\{m^b_\l\} , \{s^b_\p\}, \{ \theta^b_\v \} \rangle \right\}$ and $\left\{ |\{m^a_{\l^\star}\} , \{s^a_{\p^\star} \}, \{ \theta^b_\v \} \rangle \right\}$ respectively. However, in order to find the eigenstates for $H_a+H_b$ -- where the two terms commute and can be simultaneously diagonalized (even so after projected into the links' low energy subspaces in the small $W$ limit) -- we must take proper superpositions within either set of basis states \cite{Kitaev:1997wr, Levin:2004mi}. In particular, we need to sum over all the aforementioned gauge equivalent states that share the same $\{v^a_{\p^\star}\}, \{v^b_\p\}$ excitation configurations and belong to the same topological class $C$ or $C^\star$, with suitable phase coefficients. We find the proper combinations of all gauge equivalent states to be
\begin{widetext}
\begin{align}
|C\rangle \equiv &  \sum_{\{t^b_\v\}}  \int_{\{\theta^b_\v\}} \ \exp\left[- \sum_\v \frac{\i}{n}  (2\pi t^b+\theta^b)_\v  \left([:\frac{p\d^\star A^\star}{2\pi}:]-v^a\right)_{\p^\star=\v}\right] \ \mathsf{T}(\{t^b_{\v}\}) \ \left|\{m^{b,C\: rep}_{\l}\}, \{s^{b, C\: rep}_\p\}, \{ \theta^b_\v \} \right\rangle \label{Cdefinition}
\\
|C^\star\rangle \equiv &  \sum_{\{ t^a_{\v^\star} \}} \int_{\{\theta^a_{\v^\star}\}} \ \exp\left[-\sum_{\v^\star} \frac{\i}{n} (2\pi t^a+\theta^a)_{\v^\star} \left([:\frac{q\d A}{2\pi}:] -v^b\right)_{\p=\v^\star} \right]
\ \mathsf{T}^\star(\{t^a_{\v^\star}\}) \: \left|\{m^{a,C^\star\: rep}_{\l^\star}\}  , \{\theta^a_{\v^\star} \}, \{ s^{a, C^\star\, rep}_{\p^\star}\} \right\rangle
\label{Cstardefinition}
\end{align}
\end{widetext}
where we introduced $[:x:] \equiv x - [x]$. The state $\left|\{m^{b,C\: rep}_{\l}\}  , \{s^{b, C\: rep}_\p\}, \{ \theta^b_\v \} \right\rangle$ is a representative state in the class $C$ and $\left|\{m^{a,C^\star\: rep}_{\l^\star}\}  , \{\theta^a_{\v^\star} \}, \{ s^{a, C^\star\: rep}_{\p^\star}\} \right\rangle$  in $C^\star$ (with some given $v^b, v^a$ configurations of excitations), and $\mathsf{T}(\{t^b_{\v}\})$ and $ \mathsf{T}^\star(\{t^a_{\v^\star}\})$ are the aforementioned $0$-form $\mathbb{Z}_n$ gauge transformations specified by $\mathbb{Z}_n$ variables $\{t^b_{\v}\}$, $\{t^a_{\v^\star}\}$, the definition of which is given in details in \eqref{app:Tdefinition} and \eqref{app:Tstardefinition}. The set of states $\{|C\rangle\}$ are manifest eigenstates of $H_b$, while the set $\{|C^\star\rangle\}$ are manifest eigenstates of $H_a$, but we now show either set can actually serve as simultaneous eigenstates of both terms, i.e. eigenstates of the entire Hamiltonian. To see this, in Appendix \ref{app_solution} we show $|C\rangle$ is a linear combination of all $|C^\star\rangle$:
\begin{align}
    |C\rangle =&\sum_{C^\star}\prod_\l \mathrm{e}^{-\i \frac{2\pi  }{n} ( m^{b,C\: rep} + \frac{qA}{2\pi })_\l \cdot (  m^{a,C^\star\: rep} + \frac{pA^\star}{2\pi})_{\l^\star}} \ 
|C^\star\rangle
\end{align}
which means either the set $\{|C\rangle\}$ or the set $\{|C^\star\rangle\}$ is a choice of orthonormal basis for the $n^2$-dimensional eigensubspace, specified by the given anyon configurations $\{v^a_{\p^\star}\}$ and $\  \{v^b_\p\}$, of the full Hamiltonian. 
 
We emphasize again that the under $2\pi$ shift of $A_\l$ (and the identified $A^\star_{\l^\star}$), the explicit expression for a given $|C\rangle$ or $|C^\star\rangle$ in the original physical basis of $|b_\l\rangle$ or $|a_{\l^\star}\rangle$ must remain unchanged, even though the representative labels $\{m^{b,C\: rep}_{\l}\}  , \{s^{b, C\: rep}_\p\}$ or $\{m^{a,C^\star\: rep}_{\l^\star}\}, \{ s^{a, C^\star\: rep}_{\p^\star}\}$ would change. Clearly, the physical excitation numbers $\{ v^b_\p \}, \{v^a_{\p^\star}\}$ and the Wilson loops $L^b_{\ell_i}$ or $L^a_{\ell^\star_{\hat{\mathbf{i}}}}$ that characterize $|C\rangle$ or $|C^\star\rangle$ remain unchanged.

We finally recall that all we have solved above was $H_a+H_b$ projected in the trial subspaces $\bigotimes_\l \mathcal{T}_\l$, whilst we really should have done it in the actual low energy subspaces $\bigotimes_\l \mathcal{L}_\l$. Since the error between $\mathcal{L}_\l$ and $\mathcal{T}_\l$ is $\mathcal{O}(\sqrt{\epsilon/U})$ for each $\l$, we may view any correction as a having an extra perturbation Hamiltonian on each link $\l$. But it is well known that in the thermodynamical limit such local perturbation terms do not alter the topological physics of the toric code model \cite{Kitaev:1997wr}, and the same reasoning applies here. Moreover, to ensure that the $H_a, H_b$ couplings do not induce large mix with the higher energy subspace $\mathcal{T}_{\perp, \l}$ on each link, we need the $H_a, H_b$ to be smaller than $\omega_a, \omega_b$. The conditions are indeed the $\epsilon \ll V \ll \sqrt{\epsilon U}$ that we mentioned at the beginning of this subsection.

\section{Hall Conductivity and Fractionalized Electric Charge}
\label{sect_Hall}

In the previous section we have controllably solved the Hamiltonian, with errors bounded by the orders $\mathcal{O}(\sqrt{\epsilon/U}, \epsilon/V, V/\sqrt{\epsilon U})$. Now we investigate the electromagnetic responses, most importantly the Hall conductivity, in both the local and the global perspectives. The local perspective is more straightforward, while the global perspective settles the exactness of the quantization of the responses, which is necessary since we have errors from the use of perturbation theory. 

First we compute the expectation of the local electric charge density \eqref{charge_density}. Carrying out the calculation in Appendix \ref{app_Hallresponse}, we find, in general,
\begin{align}
   \langle p s^b_\p \rangle &= -\frac{p}{n}\frac{q(\d A)_\p}{2\pi} + \frac{p}{n}\left(v^b + [\frac{q\d A}{2\pi}]\right)_\p \ ,  \nonumber \\[.2cm]
   \langle q s^a_{\p^\star} \rangle &=  - \frac{q}{n} \frac{p(\d^\star A^\star)_{\p^\star}}{2\pi} + \frac{q}{n} \left(v^a + [\frac{p\d^\star A^\star}{2\pi}] \right)_{\p^\star} \
\end{align}
which are independent of the class $C$. 

When the background magnetic field $\d A$ is small (and so for $\d^\star A^\star$, given the identification between $A^\star$ and $A$ in Fig.~\ref{fig_lattice_A}) and when excitations $v^b, v^a$ are absent, we find the local charge density to be
\begin{align}
\langle \rho_\v \rangle = -\frac{pq}{2\pi n} \left[ (\d A)_{\p'=\v+\hat{\mathbf{x}}/2+\hat{\mathbf{y}}/2} + (\d A)_{\p=\v-\hat{\mathbf{x}}/2-\hat{\mathbf{y}}/2} \right] \ .
\end{align}
Since the Hall conductivity $2\pi\sigma_H = \delta\langle\rho\rangle/\delta (\d A)$, this means the Hall conductivity equals $-2pq/n$ as desired, at least approximately to within our perturbative error. We remark that the result applies to any geometry of the lattice, even for the lattice with boundaries. This suggests that upon adiabatically applying a uniform $A$, charge accumulates onto the boundary, another well-known manifestation of the Hall conductivity.

On the other hand, when the electromagnetic background is absent but $v^b, v^a$ excitations present, we have
\begin{align}
\langle \rho_\v \rangle = \frac{q}{n} v^a_{\p^\star=\v} + \frac{p}{n} v^b_{\p=\v-\hat{\mathbf{x}}/2-\hat{\mathbf{y}}/2} \ ,
\end{align}
which means the $v^a$ and $v^b$ anyons carry fractionalized electric charges $q/n$ and $p/n$ respectively. Another manifestation of the anyons' electric charges is the coefficients of the $A$ dependence in the $L^b$ and $L^a$ Wilson lines, \eqref{Lb_loop} and \eqref{La_loop}, which are the worldline operators that create the $v^a$ and $v^b$ anyons respectively.

Now we consider $(\d A)_\p$ being not so small, so that it is closer to an integer $w^b_\p$ multiple of $-2\pi/q$. Then we can see from \eqref{eigenvaluesb} that the ground states with $v^b=0$ already energetically prefer $m^b$ configurations that have $w^b_\p$ fluxes at $\p$. Then the Wilson loop operator $L^b_\ell$ with $\ell$ enclosing $\p$ indeed picks a phase $\mathrm{e}^{\i \frac{2\pi}{n}w^b_\p}$ besides the regular part $\mathrm{e}^{\i \frac{q}{n}(\d A)_{\p}}$, and the charge on $\p$ also receives a contribution $-\frac{p}{n}w^b_\p$. Therefore, these $\mathbb{Z}_n$ gauge fluxes should also be intepreted as flux anyons, but attached to the background fluxes. Likewise, a magnetic flux $(\d^\star A^\star)_{\p^\star}$ close to an integer $w^a_{\p^\star}$ multiple of $2\pi/p$ creates $w^a_{\p^\star}$ charge anyons at $\p^\star$ in the ground state. When some $(\d A)_\p$ or $(\d^\star A^\star)_{\p^\star}$ take $(2\pi/q)(\mathbb{Z}+1/2)$ or $(2\pi/p)(\mathbb{Z}+1/2)$ value, the Hamiltonian becomes gapless in the vicinity of $\p$ or $\p^\star$, undergoing a transition to a new ground state with different anyon numbers.

The results above are all computed using the trial wavefunctions, which differ from the actual wavefunctions with errors controlled by $\mathcal{O}(\sqrt{\epsilon/U}, \epsilon/V, V/\sqrt{\epsilon U})$. It is well-known that the $\mathbb{Z}_n$ toric code topological physics discussed in the previous section are robust against such local errors \cite{Kitaev:1997wr}. One can envision that the topological electromagnetic responses -- the Hall conductivity and the anyons' electric charges -- are robust for similar reasons. In the below we focus on the Hall conductivity. We present a global definition of Hall conductivity according to \cite{niu1985quantized} (which is also the setup employed to prove the Kapustin-Fidkowski no-go theorem \cite{kapustin2020local}), which is robust in the thermodynamical limit, and show it must be precisely quantized to our designated value $-2pq/n$ rather than any other nearby fraction blurred by the use of perturbation theory. The strategy is:
\begin{enumerate}
\item We employ the method of \cite{niu1985quantized} to show that the globally defined Hall conductivity must be an integer multiple of the inverse of the ground state degeneracy, $1/n^2$.
\item We can compute that the value is close to $-2pqn/n^2$, controlled up to errors bounded by $\mathcal{O}(\sqrt{\epsilon/U}, \epsilon/V, V/\sqrt{\epsilon U})$.
\item By adiabatically changing the parameters, we can make the errors much smaller than the spacing $1/n^2$ and eventually approach $0$ arbitrarily closely, so we can conclude the Hall conductivity must be exactly $-2pqn/n^2$; since the gap is not closed when we decreased the error adiabatically, the Hall conductivity must take this value for some finite range of parameters away from the zero error limit.
\end{enumerate}

Following \cite{niu1985quantized}, we consider a square lattice with periodic boundary conditions, forming a torus, and apply a uniform background, in which the $A_\l$ on all $x$-direction links take value $\alpha_x/N_x$ and the $A_\l$ on all $y$-direction links take value $\alpha_y/N_y$, where $N_i$ is the number of sites in the $i$-direction (the identification of $A^\star$ with $A$ is understood), so that there is no magnetic field, but only flat holonomies $(\alpha_x, \alpha_y)$. Consider the two-parameter space where $\alpha_x$ and $\alpha_y$ vary from $0$ to $2\pi n$. Suppose we fix $\alpha_y$ and vary $\alpha_x$ adiabatically (so there is a weak electric field in the $y$-direction) from $0$ to $2\pi n$. Each time $\alpha_x$ takes $2\pi\mathbb{Z}$ value, the background holonomies is gauge equivalent to the original $(\alpha_x=0, \alpha_y)$. However, while the same background is revisited $n$ times over this adiabatic process, each time the ground state (under the given background $(\alpha_x=0, \alpha_y)$) may not be same one, until it must come back to the original one at the $n$th time. This can be seen by first performing a gauge transformation of $A$ so that only one column of $x$-direction links have non-zero values, with $A_\l=\alpha_x$, and then carefully examining the definition of $|C\rangle$ and the definition of the non-contractible Wilson loops $L^b_{\ell_x}$, $L^b_{\ell_y}$ that characterize $C$. (In particular, one will find that, when $p, q$ are both coprime with $n$, the adiabatically evolving state would reach $n$ distinct ground states out of the total $n^2$, while for other values of $p, q$ not both coprime with $n$, some of the visited ground states would already be revisited before the $n$th time.) Only when $\alpha_x$ reaches $2\pi n$ are we guaranteed to return to the original state; one may note, however, that there is an overall phase that depends on $A$, and this is in fact related to the Chern number below. The same reasoning applies if we fix $\alpha_x$ and vary $\alpha_y$. Therefore, in the two-parameter space where $\alpha_x$ and $\alpha_y$ adiabatically vary from $0$ to $2\pi n$, at the $n^2$ points where $(\alpha_x, \alpha_y)$ are gauge equivalent to $(0, 0)$, the adiabatic state has visited the degenerate ground states under the $(0, 0)$ background holonomies for $n^2$ times (exhausting all the $n^2$ degenerate ground states if $p, q$ are both coprime with $n$, but otherwise missing some and repeating some).

The globally defined Hall conductivity is given by the Chern number over the $\alpha_x, \alpha_y \in [0, 2\pi n)$ space of holonomies, averaged over $n^2$ visited states \cite{niu1985quantized}:
\begin{align}
   2\pi \sigma_H &= \frac{1}{n^2} \int_{0}^{2\pi n} d\alpha_x \int_{0}^{2\pi n} d\alpha_y \ \frac{\mathcal{B}}{2\pi}, \nonumber \\[.3cm] 
   \mathcal{B} &\equiv -i \left( \left\langle\frac{\partial C_0}{\partial \alpha_x}\bigg{|}\frac{\partial C_0}{\partial \alpha_y} \right\rangle - \left\langle\frac{\partial C_0}{\partial \alpha_y}\bigg{|}\frac{\partial C_0}{\partial \alpha_x} \right\rangle \right) \ . 
\end{align}
Here $\mathcal{B}$ is the Berry curvature in the space of holonomies, and importantly the integral of $\mathcal{B}/2\pi$ is the Chern number that must be an integer. Carrying out the calculation in Appendix \ref{app_Hallresponse}, we find the Chern number is indeed $-2pqn$, and hence the Hall conductivity is indeed $-2pq/n$ as expected. (In fact, in our particular model, the Berry curvature is constant over the space of holonomies, $\mathcal{B}=-pq/\pi n$.) In the calculation the largest error is still the difference between our trial ground states and the actual ground states, bounded by $\mathcal{O}(\sqrt{\epsilon/U}, \epsilon/V, V/\sqrt{\epsilon U})$, hence our strategy above applies. The Hall conductivity must be the designated fraction within a finite range of parameters. This completes the proof of the exactness and robustness of the Hall conductivity.

\section{Conclusion}
\label{sect_conclusion}

In this paper we constructed a class of lattice Hamiltonians that can be solved at low energies and exhibit fractional Hall conductivity. The construction is systematic, motivated by the doubled Chern-Simons theory description of the associated bosonic topological orders enriched with electromagnetic $U(1)$; the solution is obtained by a combination of perturbative and exact techniques. In forthcoming works we will elaborate on generalizations towards twisted bosonic topological orders as well as the fermionic ones.

The present work is, to our knowledge, the first solution to lattice Hamiltonians that exhibit fractional Hall conductivity. (The relation to the previous literature is explained at the end of Section \ref{sect_Hamiltonian}.) Our construction method is of theoretical significance, particularly in light of the Kapustin-Fidkowski no-go theorem \cite{kapustin2020local} which forbids Hall conductivity in any local commuting projector Hamiltonian with finite dimensional local Hilbert space. Since the previously existing methods to systematically construct exactly solvable lattice Hamiltonians for topological phases are mostly subjected to these limiting conditions, our work, going beyond these constraints, may shed new light on the methodology in the more general studies of topological phases.

Let us further our discussion on the constraints in the Kapustin-Fidkowski no-go theorem. There are three non-trivial constraining assumptions in the statement of the theorem: the local Hilbert space is finite dimensional, the terms of the Hamiltonian commute, and the terms are projectors. Our construction breaks all three assumptions. It is interesting to ask whether one may break less of these constraining assumptions and still have Hall conductivity. Since commutativity is key to exact solvability, let us first consider the scenarios where commutativity is assumed. 
\begin{itemize}
\item If the local Hilbert space remains finite dimensional, and the terms of the Hamiltonian are commuting, then relaxing the projector assumption does not help because the terms can be smoothly deformed to projectors without closing the gap.
\item At the end of \cite{kapustin2020local}, the possibility of local commuting projector Hamiltonian on infinite dimensional local Hilbert space was mentioned, but such models are deemed not physical, as the Hamiltonian's matrix elements would cease to be continuous in the local dynamical and background variables on the lattice. 
\item In the same discussion, the possibility of further relaxing the projector assumption was also mentioned. To us, this possibility seems unlikely to help. In order for the terms in the Hamiltonian to be more physical than projectors, their matrix elements should be continuous functions of the local dynamical and background variables on the infinite dimensional local Hilbert space. But these terms are also assumed commuting. Then it seems the Hamiltonian will always be gapless even in finite system sizes, as is in the case of our ``prototype'' Hamiltonian $\wt{H}$.
\end{itemize}
Based on these arguments, we may conjecture that:
\begin{itemize}
\item[$\ast$] To have non-trivial Hall conductivity, it is necessary to give up the commutativity assumption. As a result, the correlations of local operators do not exactly vanish outside of any finite distance, but only decay exponentially.
\end{itemize}
(Note the conjecture is made given the aforementioned physical requirement that the Hamiltonian's matrix elements must be continuous in the dynamical and background variables on the lattice.)
Such non-commuting Hamiltonians are not exactly solvable in general. If the arguments above can be turned into a rigorous proof to the conjecture, then we would have a theorem generalizing the theorem in \cite{bezrukavnikov2019localization} from non-interacting Chern insulators to interacting systems. A natural question to ask is then:
\begin{itemize}
\item[$\ast$]
Can one construct a Hamiltonian, over finite dimensional local Hilbert space and with non-commuting terms, that is controllably solvable at lower energies, gapped, and exhibits Hall conductivity?
\end{itemize} 
(The projector assumption is no longer relevant if the terms are already non-commuting.) This is indeed a very interesting theoretical modeling problem to tackle. Moreover, limiting the local Hilbert space to be finite dimensional is usually seen as a desired ``physical'' feature, because it means the Hamiltonian can be realized in generalized ``spin'' systems at least in principle. (For instance, recently \cite{Wang:2021smv} constructed exactly solvable lattice Hamiltonians for a large class of topological orders enriched by electromagnetic $U(1)$ global symmetry but without Hall conductivity, and the Hilbert space being finite dimensional was emphasized, in comparison to the previous examples \cite{Levin:2011hq}.) We would like to remind, however, that in the conceivable proposals to realize such generalized ``spin'' models in solid state or cold atom systems, that generalized ``spin'' -- the finite dimensional local Hilbert space -- is always the local lower energy subspace of some infinite dimensional full local Hilbert space; but this situation is indeed what happens in our models. Therefore, in this sense, a theoretical model with finite dimensional local Hilbert space is not necessarily more ``physical'' than our ones with infinite dimensional local Hilbert space.

\acknowledgements
J.-Y.~C. is supported by NSFC under Grants No.12042505.

\appendix
\begin{widetext}

\section{From Effective Lagrangian to Toy Hamiltonian}
\label{app_L_to_H}

In \cite{Chen:2019mjw} it was shown that the doubled Chern-Simons theory \eqref{doubleCS} has a Lagrangian description on \emph{effective} spacetime lattice -- where by \emph{effective} spacetime lattice we mean a coarse grained spacetime manifold; the conceptual importance of this interpretation will become clear soon. For simplicity we will think of a three-dimensional cubic lattice, but the discussions below can be straightforwardly applied to any tetrahedral decomposition of a three-dimensional manifold, or a triangulation of a two-dimensional spatial manifold together with a discretization of time (hence a ``prism'' decomposition of the spacetime). The action is 
\begin{align}
S =& \ \frac{n}{2\pi} \sum_{plaq. \ p} a_p (db)_p - n\sum_{plaq. \ p} a_p s^b_p - n\sum_{link \: l} s^a_l b_l + \sum_{cube \: c} \theta^a_c (ds^b)_c + \sum_{vert. \: v} \theta^b_v (\partial s^a)_v \nonumber \\[.2cm]
& \ - \frac{q}{2\pi} \sum_{link \: l} A_l (\partial a-2\pi s^a)_l - \frac{p}{2\pi} \sum_{plaq. \: p} A^\star_p (db-2\pi s^b)_p + \sum_{plaq. \: p} a_p L^a_p + \sum_{link \: l} b_l L^b_l \ .
\label{spacetime_lattice_action}
\end{align}
Here the dynamical variables include: $a_p$ and $b_l$ take real values on plaquette $p$ and link $l$ respectively (one may view $a_p$ as living on the dual lattice, with $l^\star=p$ in three-dimensions, and likewise for other quantities with superscript $^a$), $s^a_l$ and $s^b_p$ take integer values on link $l$ and plaquette $p$ respectively, $\theta^a_c$ and $\theta^b_v$ take $U(1)=\mathbb{R}/2\pi\mathbb{Z}$ values on cube $c$ and vertex $v$ respectively. The background fields $A_l$ and $A^\star_p$ take real values on link $l$ and plaquette $p$ respectively; they may either the independent, or identified in a certain way, say, $A^\star_p=A_{l=p+\hat{\mathbf{x}}/2+\hat{\mathbf{y}}/2+\hat{\mathbf{z}}/2}$ (which can be generalized to other discretizations of the spacetime as long as there is a branching structure). The Wilson loop observables $L^a_p$ and $L^b_l$ that take integer values on plaquette $p$ and link $l$ respectively. $d$ is the lattice coboundary operator, i.e. lattice exterior derivative, and $\partial$ is the lattice boundary operator which is the inverse of $d$; one may check that
\begin{align}
\sum_{cube \: c} \theta^a_c (ds^b)_c = \sum_{plaq. \: p} (\partial \theta^a)_p s^b_p, \ \ \ \ \sum_{plaq. \: p} a_p (db)_p = \sum_{link \: l} (\partial a)_l b_l, \ \ \ \ \sum_{link \: l} s^a_l (d\theta^b)_l = \sum_{vert. \: v} (\partial s^a)_v \theta^b_v. 
\end{align}
on a lattice without boundary. We require the Wilson loop observables to be closed loops: $(\partial L^b)_v=0$, $(dL^a)_c=0$. The action has the following gauge invariances $\!\!\mod 2\pi$:
\begin{align}
& b_l \: \rightarrow \: b_l + 2\pi z^b_l + (d\varphi^b)_l, \ \ \ \ \ s^b_p \: \rightarrow \: s^b_p + (dz^b)_p, \ \ \ \ \ \theta^b_v \: \rightarrow \: \theta^b_v + n \varphi^b_v, \\[.2cm]
& a_p \: \rightarrow \: a_p + 2\pi z^a_p + (\partial\varphi^a)_p, \ \ \ \ \ s^a_l \: \rightarrow \: s^a_l + (\partial z^a)_l, \ \ \ \ \ \theta^a_c \: \rightarrow \: \theta^a_c + n \varphi^a_c,
\end{align}
where $z^a_p$ and $z^b_l$ take integer values on plaquette $p$ and link $l$ respectively, and $\varphi^a_c$ and $\varphi^b_v$ take $U(1)=\mathbb{R}/2\pi \mathbb{Z}$ values on cube $c$ and vertex $v$ respectively; note the importance of the integrity of $n$. The $z^a$ and $z^b$ parametrize $1$-form $\mathbb{Z}$ gauge invariances that reduce the $a, b$ from real valued to effectively $U(1)$ valued, and subsequently the $\varphi^a$ and $\varphi^b$ parametrize the effectively $U(1)$ ordinary ($0$-form) gauge invariances. The action also has the following invariances under changes of background variables:
\begin{align}
& A_l \: \rightarrow \: A_l + 2\pi Z_l + (d\Phi)_l, \ \ \ \ \ L^a_p \rightarrow L^a_p + q (dZ)_p, \ \ \ \ \ \theta^b_v \: \rightarrow \: \theta^b_v - q\Phi_v, \\[.2cm]
& A^\star_p \: \rightarrow \: A^\star_p + 2\pi Z^\star_p + (\partial\Phi^\star)_p, \ \ \ \ \ L^b_l \rightarrow L^b_l + p(\partial Z^\star)_l, \ \ \ \ \ \theta^a_c \: \rightarrow \: \theta^a_c - p\Phi^\star_c
\label{eff_indistinguishability}
\end{align}
where $Z_l$ and $Z^\star_p$ take integer values on link $l$ and plaquette $p$ respectively, and $\Phi_v$ and $\Phi^\star_c$ take $U(1)=\mathbb{R}/2\pi \mathbb{Z}$ values on vertex $v$ and cube $c$ respectively; note the importance of the integrity of $p, q$. The $Z$ and $Z^\star$ parametrize $1$-form $\mathbb{Z}$ gauge invariances that reduce the background $A, A^\star$ from real valued to effectively $U(1)$ valued, and subsequently the $\Phi$ and $\Phi^\star$ parametrize the effectively $U(1)$ ordinary ($0$-form) global symmetries. Apparently, if $A^\star$ is identified with $A$ in the said way, then $Z^\star, \Phi^\star$ must be identified with $Z, \Phi$ in the associated manner (but the Wilson loop observables $L^a$ and $L^b$ are still independent).

An important feature of this \emph{effective} Lagrangian theory is that the backgrounds $A_l$ and $A_l + 2\pi Z_l$ are not identical outright; an associated transformation of the Wilson loop observables is required. In physical terms, this means in this theory, a narrow thread of $2\pi$ electromagnetic flux in the background is physically indistinguishable from inserting a Wilson loop observable which creates a certain anyon worldloop. This is indeed a property of the original doubled Chern-Simons theory \eqref{doubleCS} in the continuum. A conceptually important point, however, is that this property is an effective, macroscopic one, in the sense that it only applies when the thread of the electromagnetic flux is ``narrow'' compared to the scales of interest but large compared to any microscopic scale. This is why the lattice Lagrangian theory above is only applicable to \emph{effective} spacetime lattice, by which we mean a coarse grained spacetime manifold. Were the ``lattice'' not a coarse grain but an actual \emph{microscopic} lattice, the description is inapplicable, because a $2\pi$ magnetic flux that is narrower than the microscopic lattice scale should be invisible outright, i.e. the backgrounds $A_l$ and $A_l+2\pi Z_l$ should be identical outright. It is this important difference in the physical requirements between a theory on an effective lattice (coarse grain) and one on an actually microscopic lattice that makes the usual wisdom \cite{Levin:2004mi, Kirillov:2011mk} which generates toy model Hamiltonians on microscopic lattice from coarse grained descriptions (fixed point properties) not directly applicable to our present problem \cite{Chen:2019mjw}.

Let us nonetheless proceed and obtain the Hilbert space and operator contents from this effective Lagrangian. For the cubic lattice, we view two directions as the spatial lattice and one direction as the discretized time. (We may do the same for a prism decomposition of the spacetime, where the space is triangulated and the time is discretized.) Then a three-dimensional vertex $v$ is associated with a spatial vertex $\v$ and an integer time step $t$, denoted as $v=(\v, t)$, while (the center of) a three-dimensional cube $c=(\p, t+1/2)$; on the other hand, a three-dimensional link $l$ has two possibilities $l=(\l, t)$ or $l=(\v, t+1/2)$, and a three-dimensional plaquette has two possibilities $p=(\p, t)$ or $p=(\l, t+1/2)$. In these notations, our previous action reads $S = \sum_t L_t$, with
\begin{align}
L_t \ =& \ \frac{n}{2\pi} \sum_\l a_{\l, t+1/2} \left( b_{\l, t+1} - b_{\l, t} \right) - \sum_\p \theta^a_{\p, t+1/2} \left( s^b_{\p, t+1} - s^b_{\p, t} \right)  + \sum_\v s^a_{\v, t+1/2} \left( \theta^b_{\v, t+1} - \theta^b_{\v, t} \right) \nonumber \\[.2cm]
& \ + \sum_\l s^a_{\l, t} \left[ d\theta^b - n b + qA \right]_{\l, t} + \sum_\l s^b_{\l, t+1/2} \left[ \partial \theta^a - n a + pA^\star \right]_{\l, t+1/2} \nonumber \\[.2cm]
& \ + \sum_\p \frac{a_{\p, t}}{2\pi} \left[ n \left(d b - 2\pi s^b \right) - q dA + 2\pi L^a \right]_{\p, t} + \sum_\v \frac{b_{\v, t+1/2}}{2\pi} \left[ n \left(\partial a - 2\pi s^a \right) - p dA^\star + 2\pi L^b \right]_{\v, t+1/2}
\end{align}
where for simplicity we have assumed that $A, A^\star$ only have magnetic fields but no electric fields, and $L^a$, $L^b$ only run along the time direction, i.e. the anyon insertions created by inserting these Wilson loops are held at fixed positions in the space. Only the three terms in the first line involve discretized time derivative; as usual they give rise to the local Hilbert spaces endowed with the commutation relations \eqref{ba_comm}, \eqref{sathetab_comm} and \eqref{sbthetaa_comm}, as long as we introduce the spatial dual lattice notions as explained there. On the other hand, the four terms in the last two lines all involve Lagrange multipliers, so instead of Hamiltonian terms, they give rise to strict constraints on the Hilbert space. In particular, summing over the integer valued $s^a_{\l, t}$ and $s^b_{\l, t+1/2}$ imposes the Gauss's law constraints for the $1$-form $\mathbb{Z}$ gauge transformations \eqref{1-form_Z}, generated by \eqref{generator_g}, while integrating over the real valued $a_{\p, t}$ and $b_{\v, t+1/2}$ imposes the Gauss's law constraints for the ordinary ($0$-form) gauge transformations, generated by \eqref{generator_f}:
\begin{align}
g^b_\l = e^{-iqA_\l}, \ \ \ \ \ g^a_{\l^\star} = e^{-ip A^\star_{\l^\star}} \ ; \ \ \ \ \ \ f^b_\p = \frac{q}{n} (\d A)_\p - \frac{2\pi}{n} L^a_{\v^\star=\p}, \ \ \ \ f^a_{\p^\star} = \frac{p}{n} (\d^\star A^\star)_{\p^\star} - \frac{2\pi}{n} L^b_{\v=\p^\star}.
\label{full_constraints}
\end{align}
The commutation relations along with these Gauss's law constraints is the full content of the theory; there is no Hamiltonian on top of the subspace specified these constraints.

While this theory is well-defined, there are two undesired features, given that our goal is to construct microscopic lattice (toy) model Hamiltonians:
\begin{enumerate} 
\item
Again, a $2\pi$ background magnetic flux $(\d A)_\p$ through a single plaquette $\p$ is not invisible, but only equivalent to an $L^a_\p$ insertion taking value $-q$. Therefore the ``spatial lattice'' here must be viewed as a coarse grain rather than an actual microscopic spatial lattice. We are interested in theories (albeit toy theories) on microscopic spatial lattice, in which a $2\pi$ background magnetic flux $(\d A)_\p$ through a single plaquette $\p$ is invisible outright.
\item
The current theory is a lattice gauge theory with strict Gauss's law constraints imposed on the physical Hilbert space. We are interested in theories with no strict constraints on Hilbert space, and any appearance of gauge constraint should be emergent at low energies \cite{Kitaev:1997wr, Levin:2004mi}.
\end{enumerate}
Therefore the remaining task is to modify the theory to evade these two issues.

To resolve the first issue, partly motivated by \cite{Levin:2011hq}, in $L_t$ we may strip off the direct coupling of the backgrounds $A, A^\star$ to the real valued $a, b$, so that the backgrounds only couple to integer valued variables $s^a$, $s^b$, and hence their $2\pi\mathbb{Z}$ parts indeed do not matter, and they are indeed $U(1)$; as a result of this, the $f^b, f^a$ constraints in \eqref{full_constraints} can no longer be gauge constraints, but they can emerge as energetic conditions. Consider the lattice gauge theory with two $1$-form $\mathbb{Z}$ Gauss's law constraints and a Hamiltonian:
\begin{align}
& g^b_\l = e^{-iqA_\l}, \ \ \ \ \ g^a_{\l^\star} = e^{-ip A^\star_{\l^\star}} \nonumber \\[.2cm]
& H_{gauge} = \frac{V_b}{2} \sum_\p \left(f^b_\p\right)^2 + \frac{V_a}{2} \sum_\v \left(f^a_\v\right)^2. 
\label{gauge_theory}
\end{align}
In this theory the background fields $A_\l$ and $A^\star_{\l^\star}$ are indeed $U(1)$ valued as desired. (In the above we omitted electric field for convenience; it can be shown that the theory \eqref{gauge_theory} stays the same even if include electric field via the time dependence of $A$ and $A^\star$.) Moreover, under the Gauss's law constraints, minimizing the Hamiltonian $H_{gauge}$ will indeed lead to the $f^a, f^b$ constraints in \eqref{full_constraints}, with $L^a_{\v^\star=\p}$ given by the integer closest to $q(\d A)_\p$ and $L^b_{\v=\p^\star}$ given by the integer closest to $p(\d^\star A^\star)_{\p^\star}$ -- this means large enough magnetic fluxes can create anyon insertions, and one reminiscence of this fact is the aforementioned macroscopic effective indistinguishably between a finite size $2\pi$ background magnetic flux and a certain Wilson loop insertion.

The theory \eqref{gauge_theory} is, however, still a gauge theory. To resolve this second issue, we recall that in the previous studies of exactly solvable models, gauge constraints are energetically imposed \cite{Kitaev:1997wr, Levin:2004mi}. This motivates us to view \eqref{gauge_theory} as the $U_a, U_b \rightarrow \infty$ limit of the ``prototype'' Hamiltonian $\wt{H}$ introduced in Section \ref{sect_Hamiltonian}. This is how we motivate for $\wt{H}$. Upon making $U_a, U_b$ any finite values, however, the local gaplessness problem arises, as explained in Section \ref{sect_Hamiltonian}. This problem occurs here but not in \cite{Kitaev:1997wr, Levin:2004mi} because the local operators here take continuous rather than discrete values, in order to accommodate for suitable couplings to the continuous background $U(1)$ gauge field(s) $A$ and $A^\star$. The resolution to this local gaplessness problem finally led us to our construction of the Hamiltonian $H$.

To further understand how the discussion so far is related to the Kapustin-Fidkowski no-go theorem, let us first review how, when either $p$ or $q$ vanishes (and hence the Hall conductivity vanishes), one may construct an exactly solvable lattice Hamiltonian \cite{Levin:2011hq} that resolves the two issues above. For concreteness we take $q=0$. Then, on each link $\l$, instead of the real valued operator $b_\l$ subjected to the constraint $g^b_\l=1$ in \eqref{full_constraints}, we can use a finite local Hilbert space endowed with
\begin{align}
& \wt{b}_\l \in \left\{ 0, 1, \cdots, n-1 \right\}, \ \ \ \ \ \ \left[\wt{b}_\l, e^{-i\wt{a}_{\l^\star}}\right] = e^{-i\wt{a}_{\l^\star}} \ \ \mbox{except} \ \ e^{-i\wt{a}_{\l^\star}} \left| \wt{b}_\l=n-1 \right\rangle = 0, \ \ e^{i\wt{a}_{\l^\star}} \left| \wt{b}_\l=0 \right\rangle = 0.
\end{align}
We may view $\wt{b}_\l$ as a reminiscence of $(nb-\d\theta^b)_\l/2\pi$. Since the local Hilbert space is now finite, the other constraints in \eqref{gauge_theory} can be energetically imposed by the local commuting gapped Hamiltonian
\begin{align}
H_{comm.} = \frac{V}{2} \sum_{\p} \left( \d\wt{b} - ns \right)_\p^2 + U \sum_{\v=\p^\star} \left[ \left| 1 - e^{i (\d^\star \wt{a})_{\p^\star}} \right|^2 \prod_{\l^\star\in\partial \p^\star} \left| 1 - e^{i(\d^\star\theta - n\wt{a} + pA^\star)_{l^\star}} \right|^2 \right]
\end{align}
where $s_\p, \theta_{\v^\star=\p}$ are reminiscence of $s^b_\p, \theta^a_{\v^\star=\p}$, while $s^a_{\p^\star}$ and $\theta^b_{\v=\p^\star}$ are gone; note that the $e^{\mp i n\wt{a}}_{\l^\star}$ in the last factor is non-vanishing only if it is multiplied to an $e^{\pm i \wt{a}_{\l^\star}}$ from the previous $e^{\pm i (\d^\star \wt{a})_{\p^\star}}$ factor. When $p\neq 0$ and $n\neq 1$, we may solve for fractionally charged anyons \cite{Levin:2011hq}; when $p=0$, we may further simplify the model to a $\mathbb{Z}_n$ toric code. Recently in \cite{Wang:2021smv} it was emphasized that the topological low energy physics of this model does not involve states with large values of $|s_\p|$, so the local Hilbert space on $\p$ can also be trimmed to finite, making all the local Hilbert spaces finite dimensional (in fact \cite{Wang:2021smv} encompassed more general cases including twisted bosonic topological orders, as long as the Hall conductivity vanishes). 

Apparently, such exactly solvable Hamiltonian becomes unavailable when both $p,q $ are non-zero, i.e. when the Hall conductivity does not vanish, in echo with the Kapustin-Fidkowski no-go theorem \cite{kapustin2020local}. In particular, the background fields must couple to the system only through the Hamiltonian, but not the eigenvalues of any local microscopic operator, so we may no longer define some $\wt{b}_\l$ as a reminiscence of $(nb-\d\theta^b)_\l/2\pi$ under the constraint $g^b_\l = e^{-iqA_\l}$ when $q\neq0$. On the other hand, we may not define $\wt{b}$ as a reminiscence of $(nb-\d\theta^b-qA)_\l/2\pi$ either, because in that case the first term of $H_{comm.}$ would involve $(\d\wt{b} - ns-q\d A)^2$, violating the invisibility requirement of a $2\pi$ background flux; we may not use $\cos(\d\wt{b} - ns-q\d A)$ either because then the $s$ operator which importantly carries electric charge (when $p\neq 0$) would have dropped out. This is why we emphasize \cite{Chen:2019mjw} that the usual procedure to obtain exactly solvable lattice Hamiltonian from an exactly solvable coarse grained effective theory becomes unsuccessful here for general values of couplings $p, q$, due to the difference in the physical requirements satisfied by a $2\pi$ background flux on a coarse grained ``lattice'' versus on an actual microscopic lattice.

\section{Details for Local Low Energy Subspace}
\label{app_LLES}
The local Hamiltonian on each link can be expressed as (the link index is omitted in this section):
\begin{equation} \label{app:localH}
H_{\text{link}} = \frac{1}{2}\left(\epsilon_b b^2 +\epsilon_a a^2\right) - U_b \cos[n(b-\bar{b}_0)] - U_a \cos[n(a - \bar{a}_0)]
\end{equation}
where we remind the commutation relation is $[b,a] = \i 2\pi/n$, and $\bar{a}_0$, $\bar{b}_0$ are terms that commute with $a$ and $b$; the electromagnetic background couples to the system through $\bar{a}_0, \bar{b}_0$. The task of this section is to perturbatively find the low energy subspace of $H_{\text{link}}$ under the assumption $\epsilon\ll U$ (where $\epsilon$ is the scale of $\epsilon_a$ and $\epsilon_b$, and $U$ the scale of $U_a$ and $U_b$). The low energy subspace turns out to be an emergent $\mathbb{Z}_n$ space with nearly degenerate energies. More particularly, we will:
\begin{enumerate}
\item construct $n$ trial wavefunctions that are nearly orthogonal up to error exponentially small in $\sqrt{U/\epsilon}$. Denote the space spanned by them as $\mathcal{T}$;
\item show there is large energy gap $\mathcal{O}(\sqrt{\epsilon U})$ between when $H_{\text{link}}$ is projected into $\mathcal{T}$ and into the orthogonal subspace $\mathcal{T}_\perp$;
\item show $H_{\text{link}}$ projected into $\mathcal{T}$ is nearly proportional to the identity matrix, up to error $\mathcal{O}(\epsilon)$, which is smaller by $\mathcal{O}(\sqrt{\epsilon/U})$ compared to the large gap $\mathcal{O}(\sqrt{\epsilon U})$;
\item show $H_{\text{link}}$ mixes $\mathcal{T}$ and $\mathcal{T}_\perp$ at order $\mathcal{O}(\epsilon)$, again smaller by $\mathcal{O}(\sqrt{\epsilon/U})$ compared to the large gap $\mathcal{O}(\sqrt{\epsilon U})$.
\end{enumerate}
This means $H_{\text{link}}$ has an actual low energy subspace $\mathcal{L}$ that is $n$ dimension, well separated in energy from $\mathcal{L}_\perp$, and the energy split within $\mathcal{L}$ is $\mathcal{O}(\epsilon)$ which is $\mathcal{O}(\sqrt{\epsilon/U})$ compared to the gap $\mathcal{O}(\sqrt{\epsilon U})$; moreover, the error between the actual $\mathcal{L}$ and the trial $\mathcal{T}$ is $\mathcal{O}(\sqrt{\epsilon/U})$.

To motivate our construction of the trial wavefunctions, first we only take $\epsilon_a$ and $U_b$ terms into consideration,
\begin{align}\label{app:Hlinkb}
    H_{\text{link},b} = \frac{1}{2}\epsilon_a a^2  - U_b \cos[n (b-\bar{b}_0)]
\end{align}
which analogously describe the one-dimensional, quantum mechanical motion of a particle in a sinusoidal potential. As long as the potential is deep enough, the neighborhood around each minimum, located at $\bar{b}_j = \frac{2\pi}{n}j + \bar{b}_0$ with $j\in \mathbb{Z}$, can be effectively described by the Taylor expansion to the quadratic order. Therefore, the low energy eigenfunctions are approximately those of infinitely many harmonic oscillators at the potential minima, the approximate Hamiltonians of which can be written as
\begin{align}\label{app:ladderc}
    H_{j} &\approx \frac{1}{2}\epsilon_a a^2 + \frac{1}{2} U_b n^2 (b - \bar{b}_j)^2= \omega_b\left(c_j^\dagger c_j +\frac{1}{2}\right)
\end{align}
with excitation energy
\begin{align}
    \omega_b = 2\pi \sqrt{U_b\epsilon_a}
\end{align}
and standard construction of ladder operators $c_j$.
The solution of the low energy states are:
\begin{align}
    |\phi^{(N_b)}_{j}\rangle \equiv  \int d b ~ C^{(N_b)} H^{(N_b)}\left(\frac{b-\bar{b}_j}{W_b}\right)\mathrm{e}^{-(b-\bar{b}_j)^2/2W_b^2}|b\rangle
\end{align}
where $N_b$ represent the energy level, $C^{(N_b)}=\frac{1}{\pi^{1/4} \sqrt{W_b} \sqrt{2^{N_b} N_b!}}$ is a normalization factor, $H^{(N_b)}(x)$ is the $N_b$-th order Hermite polynomial, and \begin{align}
    W_b = \frac{\sqrt{2\pi}}{n} \left(\frac{\epsilon_a}{U_b}\right)^{\frac{1}{4}}
\end{align}
is the half width of each state. For finding the low energy subspace we only need $N_b=0$, and for controlling errors we will only include small values of $N_b$ in later considerations.

To justify our negligence of the higher order terms in the Taylor expansion of cosine function at each minimum, and our elimination of the overlap between different minimums' orbitals, we will need the width of each state to be much smaller than the distance between two minima of the potential. We thus reach our first ``separation of length scale'' condition of the solvable limit:
\begin{align} \label{app:lengthcondition1}
W_b \ll \frac{2\pi}{n} \implies \frac{\epsilon_a}{U_b} \ll (2\pi)^2 \ .
\end{align}
Near this limit, we can estimate the correction of a small but finite $W_b$ to the energy gap. Firstly, the higher order of the expansion of the potential $U_b [1- \cos[n (b-\bar{b}_0)]] \approx U_b [\frac{n^2(b-\bar{b}_j)^2}{2!} - \frac{n^4 (b-\bar{b}_j)^4}{4!}+\dots]$ lead to the correction of energy gap between the ground states and the first excited states:
\begin{align} \label{app:Deltax}
    \Delta_b  &\equiv  \langle \phi^{(1)}_{j}|H_{\text{link},b}|\phi^{(1)}_{j}\rangle - \langle \phi^{(0)}_{j}|H_{\text{link},b}|\phi^{(0)}_{j}\rangle\approx \omega_b \left\{ 1- \frac{n^2}{6}W_b^2 + \mathcal{O}\left(W_b^4\right)\right\}
\end{align}
(same for all $j$). This correction decreases the gap while it indeed approaches zero as $W_b \rightarrow 0$. There is also an order $\mathcal{O}(W_b^2)$ mixing between different $N_b$ states. Secondly, the overlap integral between neighboring orbits 
\begin{align}\label{app:tx}
t_b \equiv \langle \phi^{(0)}_{j}|H_{\text{link},b}|\phi^{(0)}_{j+1}\rangle \approx \omega_b \frac{\pi^2}{n^2W^2_b} \mathrm{e}^{-\pi^2/n^2W_b^2}
\end{align}
can lead to an energy splitting up to $4t_b$ in the infinitely degenerate ground states space. Since $t_b$ is exponentially suppressed, compared to other quantities of order $\mathcal{O}(1)$ or other errors of order $\mathcal{O}(W_b^2)$, we can safely neglect the splitting in the said limit.

Now we have a new set of basis states describing the low energy physics of the system, spanned by infinitely many highly localized orbits $|\phi_j^{(N_b)}\rangle$. Our next step would be solving the remaining terms,
\begin{align}
    H_{\text{link},a} = \frac{1}{2}\epsilon_b  b^2 - U_a \cos[n (a-\bar{a}_0)]
\end{align}
in this basis. Note that the $U_a$ term commutes with $H_{\text{link}, b}$ and only changes $|\phi_j^{(N_b)}\rangle$ to $|\phi_{j+1}^{(N_b)}\rangle$. It remains to consider the $\epsilon_b$ term in the $|\phi_j^{(N_b)}$ basis. We approximate
\begin{align} \label{app:neglect}
    b^2 |\phi_j^{(N_b)} \rangle  = \bar{b}_j^2 |\phi_j^{(N_b)} \rangle + (\cdots),
\end{align}
where the $\bar{b}_j \equiv (2\pi/n) \hat{j} + \bar{b}_0$ operator (with $\hat{j}$ measuring the value of $j$), well-defined at least for states $|\phi^{(N_b)}_{j}\rangle$ with small values of $N_b$, measures the center coordinates. We claim that for small values of $N_b$ the neglected terms $(\cdots)$ are suppressed under suitable assumptions of parameters; we will justify the claim in details later. This means that at low energies we no longer need to consider the continuous $b$ variable but can concentrate on a discrete $\bar{b}$ space for each level $N_b$ (only small $N_b$ are of interest). The basis of the conjugate variable of $\bar{b}$ operator, $\underline{a}$ (not to be confused with $\bar{a}$), is then defined by 
\begin{align}
    |\underline{a}^{(N_b)}\rangle \equiv \sum_j \mathrm{e}^{\i  \bar{b}_j \underline{a}} |\phi^{(N_b)}_j\rangle
\end{align} 
and is the projection of $a$. The value of $\underline{a}$ is confined onto the `first Brillioun zone' in the reciprocal space of the $\bar{b}_j$ lattice, i.e. $\underline{a} \in [0,2\pi)$. Since the roles of $\underline{a}$ and $\bar{b}$ are conjugate, we can switch the perspective and now interpret $\underline{a}$ as a coordinate on a ring. Thus we can regard $-U_a \cos[n (\underline{a}-\bar{a}_0)]$ term as a sinusoidal potential subjected to the periodic boundary condition identifying $\underline{a}=0$ and $\underline{a}=2\pi$. This potential has $n$ minima located at $\bar{a}_{m^a} = \frac{2\pi m^a}{n} +\bar{a}_0$, $m^a\in \mathbb{Z}_n$. We can solve the Hamiltonian around each minimum, again treating the potential in the neighbors of the minima as quadratic:
\begin{align}
    H_{m^a} &\approx \frac{1}{2}\epsilon_b \bar{b}^2 +\frac{1}{2} U_a  n^2 (\underline{a}-\bar{a}_{m^a})^2 = \omega_a \left(d_{m^a}^\dagger d_{m^a} +\frac{1}{2} \right)
\end{align}
with the standard construction of ladder operators $d_{m}$, and excitation energy 
\begin{align}
    \omega_a = 2\pi \sqrt{\epsilon_b U_a} .
\end{align}
Then the eigenstates under this approximation are:
\begin{align}
    |\tilde{\Psi}^{(N_b,N_a)}_{m^a}\rangle= &\int \frac{\d\underline{a}}{\sqrt{2\pi}}
    C^{(N_a)}H_k\left(\frac{\underline{a}-\bar{a}_{m^a}}{W_a}\right) \mathrm{e}^{-(\underline{a}-\bar{a}_{m^a})^2/2W_a^2} |\underline{a}^{(N_b)}\rangle
\end{align}
with half width $W_a$ in the $a$ space:
\begin{align}
    W_a = \frac{\sqrt{2\pi}}{n} \left(\frac{\epsilon_b}{U_a} \right)^{\frac{1}{4}}.
\end{align}
Each level labeled by $(N_b,N_a)$ then would be $n$-fold degenerate, with energy $E^{(N_b,N_a)} = \omega_b(N_b+\frac{1}{2})+ \omega_a(N_a+\frac{1}{2})$. We will refer to the $n$ states with $N_b=N_a=0$ and $m^a=0, 1, \cdots, n-1$ as the trial ground states.

We again need the width of those states to be much smaller than the distance between two minima of the potential, and thus reach our second ``separation of length scale'' condition of the solvable limit:
\begin{align}\label{app:lengthcondition2}
   W_a \ll \frac{2\pi}{n}  \implies \frac{\epsilon_b}{U_a} \ll (2\pi)^2 
\end{align}
We can also discuss the effect of a small but non-vanishing $W_a$. Similar to the previous discussion, the higher order of the expansion of the potential lead to the correction of energy gap between the ground states and the first excitation states:
\begin{align}\label{app:Deltap}
    \Delta_a = \omega_a \left\{ 1- \frac{n^2}{6}W_a^2 + \mathcal{O}\left(W_a^4\right)\right\}
\end{align}
which is unimportant in the $W_a\rightarrow 0$ limit. Again different $N_a$ states have an order $\mathcal{O}(W_a)^2$ mixing. And the overlap integral between two neighboring orbits
\begin{align}\label{app:tp}
t_a \equiv \langle \tilde\Psi^{(0,0)}_{m^a}|H_{\text{link},a}| \tilde\Psi^{(0,0)}_{m^a+1}\rangle \approx \omega_a \frac{\pi^2}{n^2W_a^2} \mathrm{e}^{-\pi^2/n^2W_a^2}
\end{align}
can lead to an exponentially small energy splitting of up to $4t_a$ among the $n$-fold degenerate ground states, which can be neglected in the $W_a\rightarrow 0$ limit. 

Now we come back to the approximation $b^2\rightarrow \bar{b}^2$ in \eqref{app:neglect} that is yet to be justified. We show this can be justified in under yet another condition. Writing $b^2=\bar{b}_j^2 + 2\bar{b}_j(b-\bar{b}_j) + (b-\bar{b}_j)^2 $, the terms neglected in \eqref{app:neglect} are:
\begin{align}
    \frac{1}{2}\epsilon_b (b^2 - \bar{b}^2) 
    = \omega_a \sum_{j,m^a}\i W_bW_a\frac{n}{2\pi} (c_j^\dagger + c_j) (d_{m^a}^\dagger-d_{m^a}) + \omega_a \sum_j \frac{W_a^2W_b^2}{2}\left(\frac{n}{2\pi}\right)^2 (c_j^\dagger +  c_j)^2
\end{align}
where, as said before, the $\bar{b}$ operator and the associated $d_{m^a}, d_{m^a}^\dagger$ are well defined for small values of $N_b$, and we have neglected the terms that are exponentially suppressed in $1/W_b^2$, since we are justifying the error being polynomial in $W_b^2$ or smaller. Note that although we used multiple ladder operators to express the original operator, it should be kept in mind that the problem we are considering here is just a single-body problem. It is thus remarkable that these terms cannot mix different states in the degenerate ground state manifold spanned by $\tilde{\Psi}^{(N_b,N_a)}_{m^a}$, so they can at most modify the size of the excitation gaps and the form of the trial ground states. For $N_b$ and $N_a$ of order $1$, the matrix elements of $c, c^\dagger$ and $d, d^\dagger$ are of order $1$. There are several types of terms: First, the $c c$, $c d$, $ c^\dagger d$, $c d^\dagger$ terms annhilate the trial ground states. Second, the $c^\dagger c$ terms only positively modify the excitation gap $\Delta_b$ (by the same amount for all $j$s), and hence unimportant. Third, the $c^\dagger d^\dagger$ mix the trial ground states $(N_b, N_a)=(0, 0)$ with $(1,1)$; but since $W_a, W_b$ are small, the matrix element of order $\omega_a W_b W_a$ is indeed small compared to the energy difference $\omega_b + \omega_a$ between $(0, 0)$ and $(1, 1)$. Finally, there is a $c^\dagger c^\dagger$ term with matrix element $\omega_a W_a^2W_b^2 (n/2\pi)^2/2$; to justify our approximation this term must be small compared to the energy difference $2\omega_b$ between $(0,0)$ and $(2,0)$. This leads to the last ``separation of energy scale'' condition of the solvable limit,
\begin{align}\label{app:energycondition}
 \omega_a \frac{W_a^2W_b^2}{2} \left(\frac{n}{2\pi}\right)^2 \ll 2\omega_b \implies \frac{\epsilon_b}{U_b} \ll 4 n^2 \ .
\end{align}

In summary, in a solvable limit where the three conditions \eqref{app:lengthcondition1}, \eqref{app:lengthcondition2}\&\eqref{app:energycondition} are all satisfied, we can perturbatively solve for the nearly degenerate ground states of the link Hamiltonian \eqref{app:localH}. Since we have four independent parameters, this limit is always reachable. We find the $n$-fold trial ground states 
\begin{align}\label{app:gslinktransformed} 
    |\tilde{\Psi}^{(0,0)}_{m^a}\rangle  &=   \sqrt{\frac{W_a}{\pi W_b} } \sum_{j\in\mathbb{Z}} \mathrm{e}^{\i\frac{n}{2\pi}\bar{b}_j \bar{a}_{m^a} } \mathrm{e}^{-\bar{b}_{j}^2 (n/2\pi)^2 W_a^2/2}  \int db \ \mathrm{e}^{-(b-\bar{b}_{j})^2/2W_b^2} |b \rangle     
\end{align}
with $\bar{b}_j = (2\pi/n) j + \bar{b}_0$ and $a_{m^a} = (2\pi/n)m^a+\bar{a}_0, \ m^a \in \mathbb{Z}_n$; note that when $m^a$ shifts by $n$, the wavefunction changes by an overall phase that depends on $\bar{b}_0$, which commutes with the link operators $b, a$. We can also take the linear combination that is more localized in the $b$ basis:
\begin{align}\label{app:gslink}
|{\Psi}^{(0,0)}_{m^b}\rangle &\equiv \frac{1}{\sqrt{n}}\sum_{m^a=0}^{n-1} \mathrm{e}^{-\i\frac{n}{2\pi}\bar{a}_{m^a}\bar{b}_{m^b}} |\Psi^{(0,0)}_{m^a}\rangle \nonumber \\
    &=   \sqrt{\frac{W_a}{n\pi W_b} } \sum_{z^b \in \mathbb{Z}} \mathrm{e}^{\i n z^b \bar{a}_0 } \mathrm{e}^{-\bar{b}_{m^b+nz^b}^2 (n/2\pi)^2 W_a^2/2}  \int db \ \mathrm{e}^{-\left(b-\bar{b}_{m^b+nz^b}\right)^2/2W_b^2} |b \rangle
\end{align}
Again, $m_b$ is $\mathbb{Z}_n$ valued in the sense that, when $m_b$ shifts by $n$, the wavefunction changes by an overall phase that depends on $\bar{a}_0$, which commutes with the link operators $b, a$. With two small widths, $W_b = \frac{\sqrt{2\pi}}{n}(\frac{\epsilon_a}{U_b})^{\frac{1}{4}}$, $W_a = \frac{\sqrt{2\pi}}{n}(\frac{\epsilon_b}{U_a})^{\frac{1}{4}}$, we can separate neighboring Gaussian wave packets in either $a$ or $b$ basis. Those Gaussian envelopes are very narrow and resemble delta function in one space but are very broad in the other basis. Illustrations of ${\Psi}^{(0,0)}_{m^b}$ in $b$ basis can be found in Fig.~\ref{fig_gsonlink}. Denoting the $n$-dimensional subspace spanned by the trial ground state wavefunctions as $\mathcal{T}$, the true $n$-fold ground subspace $\mathcal{L}$ of $H_{\text{link}}$ differs from $\mathcal{T}$ by a bounded error $\mathcal{O}(W_b^2, W_a^2,W_a W_b)$, and the split among the actual ground states in $\mathcal{L}$ are bounded by the same order compared to the gap with $\mathcal{L}_\perp$, which is of order $\mathrm{min}(\omega_a, \omega_b)$.

Note that we can also express the trial ground states in the $a$ basis:
\begin{align}\label{app:gslinktransformedina}
      |\tilde{\Psi}^{(0,0)}_{m^a}\rangle &=  \sqrt{\frac{W_b}{n\pi W_a} } \sum_{z^a\in \mathbb{Z}}  \int da \ \mathrm{e}^{-\i\frac{n}{2\pi} \bar{b}_0 (a-\bar{a}_{m^a})}\mathrm{e}^{-W_b^2 (n/2\pi)^2 a^2/2} \mathrm{e}^{-\left(a-\bar{a}_{m^a+nz^a}\right)^2/2W_a^2}  |a \rangle 
\end{align}
and
\begin{align}\label{app:gslinkina}
    |{\Psi}^{(0,0)}_{m^b}\rangle &=  \sqrt{\frac{W_b}{\pi W_a} } \sum_{j\in\mathbb{Z}} \int da \ \mathrm{e}^{-\i\frac{n}{2\pi} \bar{b}_{m^b} a}  \mathrm{e}^{-W_b^2 (n/2\pi)^2 a^2/2} \mathrm{e}^{-(a-\bar{a}_{j})^2/2W_a^2}  |a \rangle \ .
\end{align}
Note that these expressions are not entirely symmetric with those in $b$ basis. This is because we started with $H_{\text{link}, b}$ and only separated the excitation energy in $H_{\text{link}, b}$ from the perturbation strength. If we also impose a constraint similar to \eqref{app:energycondition}, $\epsilon_a/U_a \ll 4n^2$, we would be able to neglect the difference between those different forms and conveniently write the trial ground states in the form of \eqref{app:gslink} or \eqref{app:gslinktransformedina} for both $\Psi_{m^b}$ in $b$ basis and $\tilde{\Psi}_{m^a}$ in $a$ basis. Let's simplify the discussion by taking
\begin{align}
    \epsilon_b = \epsilon_a = \epsilon&, \ \  U_a = U_b = U, \ \ 
   W\equiv \frac{\sqrt{2\pi}}{n}\left(\frac{\epsilon}{U}\right)^{\frac{1}{4}}  \rightarrow 0
\end{align}
while fixing $\Delta_0 =2\pi \sqrt{U\epsilon}$. The asymmetry between the trial wavefunction expressions in the $b$ basis and the $a$ basis is of order $\mathcal{O}(W^2)$ (where $W$ is the width of both types of Gaussian pockets), which is indeed the order that our trial ground subspace differs from the actual ground subspace, and therefore there is no contradiction with apparent symmetry between $a$ and $b$.

\section{Justification for Projecting Many-Body Coupling Terms into Local Low Energy Subspace}
\label{app_V_terms_control}

After obtaining the local low energy subspace for $H_\text{link}$ on each link, our next step of solving the entire Hamiltonian is to project the $V_a$ and $V_b$ terms into the nearly degenerate subspace formed by tensoring these local low energy spaces. At first sight this seems problematic, because the $V_a, V_b$ terms, proportional to $(\d^\star a - 2\pi s^a)^2_{\p^\star}$ and $(\d b - 2\pi s^b)^2_{\p}$ respectively, are unbounded, and hence it seems the mixing they cause between the local low energy states and the local excited states (the gaps are of or greater than order $\sqrt{\epsilon U}$) would be unboundedly large even if we assumed $V\ll \sqrt{\epsilon U}$. Now we show this is not the case.

First of all, the remaining steps to solve for the low energy many-body topological physics does not rely on the $V_a, V_b$ terms taking this particular quadractic expression. They can be modified into other functions $F\left[(\d^\star a - 2\pi s^a)^2_{\p^\star}\right]$ and $F\left[(\d b - 2\pi s^b)^2_{\p}\right]$, where $F$ is a monotonically increasing function but bounded above when the argument is large, e.g. $F[x]=\xi[1-\exp(-x/\xi)]$ with arbitrary positive $\xi$. Hence the potential problem is immediately circumvented.

Even with our current choice of the quadratic expression, we can show the undesired mixing is indeed controlled by $V/\sqrt{\epsilon U}\ll 1$ as desired. It suffices to focus on the $V_b$ term, since any argument would work for the $V_a$ term in a similar manner. To prove this claim, we first note that the sufficient condition of treating $V_b$ term as a perturbation (compared to the energy scales in the link local Hilbert space) is to require that, for any excited state $|\Psi_k\rangle$ and any low energy state $|\Psi_0\rangle$ of local Hamiltonian $\sum_\l  H_{\text{link},\l}$  to satisfy
\begin{align} \label{app:pertubationcondition}
    \bigg{|}\langle \Psi_k | V_b(\d b - 2\pi s^a)^2_{\p} | \Psi_0\rangle \bigg{|} \ll |E_k-E_0|.
\end{align}
We now take a specific $ |\Psi_0\rangle=|\{m^b_\l\} , \{s^b_\p\}, \{ \theta^b_\v \}\rangle$, whose explicit expression is given in Eq.~\ref{bbaseexpansion}. Applying the $V_b$ term yields
\begin{align}
&(\d b - 2\pi s^b)^2_{\p} \ |\{m^b_\l\} , \{s^b_\p\}, \{ \theta^b_\v \} \rangle \nonumber\\
=& \sum_{\{z^b_\l\}} \int_{\{b_\l\}} \left[\d (b -\bar{b}_{m^b,z^b})+ \frac{2\pi}{n}([:\frac{q\d A}{2\pi}:]-v^b) \right]^2_\p \nonumber\\ & \ \ \ \ \ \  \ \  \ \ \ \ \ \ \ \ \ \ \ \ \ \  \left(\prod_\l
\ \mathrm{e}^{\i z^b_\l \cdot (p A^\star)_{\l^\star}} \mathrm{e}^{-\bar{b}_{m^b_\l,z^b_\l}^2(n/2\pi)^2W_a^2/2}  \ \mathrm{e}^{-(b_\l-\bar{b}_{m^b_\l,z^b_\l})^2/2W_b^2} \right)\bigotimes_{\l,\p,\v} |b_{\l}\rangle  |(s^b + \d z^b)_\p \rangle | \theta^b_\v \rangle \ 
\end{align} 
where $v^b$ is the number of anyonic excitaitons defined in Eq.~\ref{excitation_b}, and $[:x:]\equiv x-[x]$ always has magnitude that is less than $1/2$. Then we recognize that, up to an exponentially small error, $b_\l-\bar{b}_{m^b_\l,z^b_\l} = \frac{2\pi}{n} \frac{W_b}{
\sqrt{2}} (c^\dagger_{\l, m^b_\l+nz^b_\l}+c_{\l, m^b_\l+nz^b_\l})$ with $c^\dagger_{\l, j}$ the ladder operators introduced in Eq.~\ref{app:ladderc} on link $\l$. Therefore, the effect of $V_b$ operator on this state can be described as:
\begin{align}
    (\d b - 2\pi s^b)^2_{\p} \ \rightarrow \ \left(\frac{2\pi}{n}\right)^2\left[\frac{W_b}{\sqrt{2}}\d (c^\dagger+c) + \left([:\frac{q\d A}{2\pi}:]-v^b\right)\right]^2_\p
\end{align}
where $c^\dagger_\l \equiv \sum_j c^\dagger_{\l, j}$ change the excitation number $N_b$ of $H_{\text{link,b}}$ at $\l$ (defined in Eq.~\ref{app:Hlinkb}) by $1$ so that the energy is changed by $\omega_b$. Since $([:\frac{q\d A}{2\pi}:]-v^b)_\p$ will be an $\mathcal{O}(1)$ number for the final solution we are interested in, the amplitude of perturbing this $|\Psi_0\rangle$ out of the low energy subspace of $\sum_\l  H_{\text{link},\l}$ will be $\sim V_b W_b \sim V_b (\epsilon_a/U_b)^{1/4} $, which is small compared to the local energy gap $\omega_b \sim \sqrt{\epsilon_a U_b}$. The above description also suggests that, the eigenvalues of $V_b$ term has a systematic error $2V_b W_b^2(2\pi/n)^2$ originating from $c_\l c_\l^\dagger$ terms, relative to the expression in Eq.~\ref{eigenvaluesb}. This constant shift of energy doesn't affect any discussion in the main text.

\section{Details for Many-Body Ground States and Excitations}
\label{app_solution}

Following the solving procedure towards the many-body problem in Section \ref{sect_solution}, we need to consider the equivalent configurations generated by the $\mathbb{Z}/n\mathbb{Z}$ reminiscence of the $0$-form $\mathbb{R}/2\pi\mathbb{Z}$ gauge invariance, \eqref{0-form_U1} rescaled by $n/2\pi$ (combined with \eqref{1-form_Z}), which do not alter the conditions \eqref{excitation_b} and \eqref{excitation_a} with given values of $v^b$ and $v^a$. The explicit transformations, labeled by $t^b_\v\in\mathbb{Z}_n$ and $t^a_{\v^\star}\in\mathbb{Z}_n$ respectively, are
\begin{align}
\mathsf{T}(\{t^b_{\v}\}) |\{m^b_\l\} , \{s^b_\p
\}, \{ \theta^b_\v  \} \rangle 
&\equiv \mathrm{e}^{-\i \sum_{\l}  \lfloor \frac{m^b+\d t^b}{n}\rfloor_\l \cdot (pA^\star)_{\l^\star}} \left| \left\{ n\lfloor: \frac{m^b+\d t^b}{n}:\rfloor_\l \right\}  , \left\{\left(s^b - \d \lfloor \frac{m^b+\d t^b}{n}\rfloor\right)_\p\right\}, \{ \theta^b_\v \} \right\rangle \label{app:Tdefinition}
\\[.2cm]
\mathsf{T}^\star(\{t^a_{\v^\star}\})|\{m^a_{\l^\star}\} , \{\theta^a_{\v^\star} 
\}, \{ s^a_{\p^\star}  \} \rangle  
&\equiv \mathrm{e}^{\i \sum_{\l} (qA)_{\l} \cdot  \lfloor \frac{ m^a+ \d^\star t^a}{n}\rfloor_{\l^\star}  } \left| \left\{ n\lfloor: \frac{ m^a+ \d^\star t^a}{n} :\rfloor_{\l^\star} \right\}  , \{\theta^a_{\v^\star}\}, \left\{ \left(s^a - \d^\star \lfloor  \frac{ m^a+ \d^\star t^a}{n} \rfloor\right)_{\p^\star}\right\} \right\rangle
\label{app:Tstardefinition}
\end{align}
where we define $\lfloor x\rfloor$ to be the nearest integer that is not larger than $x$, and $\lfloor:x:\rfloor \equiv x-\lfloor x\rfloor$. For integer $x$, $n\lfloor:\frac{x}{n}:\rfloor$ simply means $x \mod n$ in the range $\{0, 1, \cdots, n-1\}$. Formally, these transformations are equivalent to substituting $\{\theta\}\rightarrow \{\theta + 2\pi t \}$. We claim that the suitable linear combinations of all gauge equivalent states that simultaneously diagonalize $H_a$ and $H_b$ are $|C\rangle$ and $|C^\star\rangle$ given in Eq.~\ref{Cdefinition}\&\ref{Cstardefinition}. These states are manifestly the eigenstates of $H_b$ and $H_a$ respectively, but we need to show that they are actually simultaneous eigenstates of both terms. To verify this, we show $|C\rangle$ is a linear combination of all $|C^\star\rangle$. We express $|C\rangle$ in the $m_a$ basis:
\begin{align}
|C\rangle = & \sum_{\{t^b_{\v}\},\{m^a_{\l^\star}\}, \{s^a_{\p^\star}\}} \prod_{\l,\v,\p} \int  d\theta^b_\v \, d\theta^a_{\v^\star} \mathrm{e}^{\i\left(s^{b, C\: rep}-\d \lfloor \frac{m^{b,C\: rep}+\d t^b}{n}\rfloor\right)_\p\cdot\theta^a_{\v^\star}}\mathrm{e}^{-\i\theta^b_{\v}\cdot (s^a)_{\p^\star}}  \nonumber\\
&\ \ \ \ \ \ \ \ \ \ \ \ \ \ \ \  
\exp\left\{-\i \frac{2\pi}{n} \left(n \lfloor:\frac{ m^{b,C\: rep} + \d t^b}{n}:\rfloor+ \frac{\d\theta^b+qA}{2\pi} \right)_\l \cdot \left( m^a +\frac{\d^\star \theta^a+pA^\star}{2\pi} \right)_{\l^\star} \right\} \nonumber\\
&\ \ \ \ \ \ \ \ \ \ \ \ \ \ \ \ 
\exp\left\{\frac{-\i (2\pi t^b_\v+\theta^b_\v) \cdot \left([:\frac{p\d^\star A^\star}{2\pi}:]-v^a\right)_{\p^\star} }{n} - \i \lfloor \frac{m^{b,C\: rep}+\d t^b}{n}\rfloor_\l \cdot (pA^\star)_{\l^\star}  \right\} |\{m^a_{\l^\star}\} , \{\theta^a_{\v^\star}
\}, \{ s^a_{\p^\star} \} \rangle \ .
\end{align}
We first treat the summation over $t^b_{\v} \in \mathbb{Z}_n$ on each vertex $\v$
\begin{align}
    \sum_{t^b_{\v}} \exp\left\{\i\frac{2\pi}{n} t^b_\v\cdot (\d^\star m^a+[\frac{p\d^\star A^\star}{2\pi}]+v^a)_{\p^\star}\right\}
\end{align}
which yields the constraint $(1/n)(\d^\star m^a+[\frac{p\phi^\star}{2\pi}]-v^a)_{\p^\star}\in \mathbb{Z}$ on each dual plaquette. Next we do the integral over $ \theta^b_\v \in (-\pi, \pi]$ on each vertex $\v$
\begin{align}
    \int d\theta^b_\v \exp\left\{ \i\theta^b_\v\cdot \left(\frac{(\d^\star m^a+[\frac{p\d^\star A^\star}{2\pi}]+v^a)_{\p^\star}}{n}-s^a_{\p^\star}\right) \right\}
\end{align}
which complete the constraint \eqref{excitation_a}. Then we can rewrite the summation over $\{m^a_{\l^\star}\}$ and $\{s^a_{\p^\star}\}$, as a combined summation over both topological class $C^\star$ and the gauge transformations $\{t^a_{\v^\star}\}$ on the representative configuration $\{m^{a,C^\star\: rep}_{\l^\star} \}$, and $\{s^{a, C^\star\: rep}_{\p^\star} \}$. We thus reach the linear combination
\begin{align}
    |C\rangle =&\sum_{C^\star}\prod_\l \mathrm{e}^{-\i \frac{2\pi}{n} ( m^{b,C\: rep} + \frac{qA}{2\pi })_\l \cdot (  m^{a,C^\star\: rep} + \frac{pA^\star}{2\pi})_{\l^\star}} |C^\star\rangle
\end{align}
as desired.

\section{Details for Hall Conductivity and Fractionalized Electric Charge}
\label{app_Hallresponse}

We first calculate the expectation value of the local electric charge density \eqref{charge_density}. In the presence of a background $A$ and arbitrary anyon excitations. assembling terms in \eqref{bbasedefinition}, \eqref{app:Tdefinition}\&\eqref{Cdefinition}, we have
\begin{align}
   \langle C |  p s^b_\p | C  \rangle 
   &= \frac{1}{\mathcal{N}}\sum_{\{t^b_{\v}\},\{z^b_{\l}\}} \prod_{\v,\l} \int_{-\pi}^{\pi}  d\theta^b_\v \ p\left(s^{b, C\: rep}+\d z^b-\d \lfloor \frac{m^{b,C\: rep}+\d t^b}{n}\rfloor\right)_\p 
   \mathrm{e}^{-\bar{b}_{n\lfloor: 
   \frac{m^{b,C\: rep} + \d t^b}{n}:\rfloor _\l, z^b_\l}^2 (n/2\pi)^2 W^2} 
\end{align}
where we formally recovered a normalization factor $\mathcal{N}$. The integral over $\{\theta_\v\}$ along with the summation over $\{t^b_{\v}\},\{z^b_{\l}\}$ is equivalent to a infinite integral over the $\bar{b}$ variable on every link, as long as each term of the integrand only depends on $\bar{b}$ variable on one link (a similar and detailed derivation of this can be found below in the calculation of the Berry curvature). Therefore, the expectation can be evaluated as:
\begin{align}
   \langle C |  p s^b_\p | C  \rangle &= \frac{1}{\mathcal{N}} \prod_{\l} \int  d\bar{b}_\l \cdot \frac{p}{n}\left(\d\bar{b}+n s^{b, C\: rep}-\d m^{b,C\: rep} - \frac{q\d A}{2\pi}\right)_\p 
   \mathrm{e}^{-\bar{b}^2 (n/2\pi)^2 W^2} \nonumber
   \\ &= -\frac{p}{n}\frac{q(\d A)_\p}{2\pi} + \frac{p}{n}\left(v^b + [\frac{q\d A}{2\pi}]\right) _\p.
\end{align}
A similar derivation gives the response on dual plaquette $\p^\star_0$:
\begin{align}
   \langle C | q s^a_{\p^\star} | C \rangle &=  - \frac{q}{n} \frac{p(\d^\star A^\star)_{\p^\star}}{2\pi} + \frac{q}{n} \left(v^a + [\frac{p\d^\star A^\star}{2\pi}] \right)_{\p^\star} \ .
\end{align}
The results are interpreted in the main text in Section \ref{sect_Hall}.

Then we calculate the globally defined Hall conductivity. It is given by the Chern number over the $\alpha_x, \alpha_y \in [0, 2\pi n)$ space of holonomies, averaged over $n^2$ visited states \cite{niu1985quantized}:
\begin{align}
    2\pi \sigma_H &= \frac{1}{n^2} \int_{0}^{2\pi n} d\alpha_x \int_{0}^{2\pi n} d\alpha_y \ \frac{\mathcal{B}}{2\pi}, \ \ \ \ \ \ \ \mathcal{B} \equiv -i \left( \left\langle\frac{\partial C_0}{\partial \alpha_x}\bigg{|}\frac{\partial C_0}{\partial \alpha_y} \right\rangle - \left\langle\frac{\partial C_0}{\partial \alpha_y}\bigg{|}\frac{\partial C_0}{\partial \alpha_x} \right\rangle \right) \ . \label{app:Chern}
\end{align}
Importantly, the integrand is a Berry curvature $\mathcal{B}/2\pi$ and the integral is the Chern number that must be an integer. Now we evaluate this integrand. For simplicity we start with the class $C_0$ in which $m^{b,C\, rep}_{\l}=0$, $s^{b, C\: rep}_{\p}=0$. First note the terms that give rise to the Berry curvature come from an average over $\langle\frac{\partial C_0}{\partial A^\star_{\l^\star}}\big{|}\frac{\partial C_0}{\partial A_{\l}}\rangle$ (or its conjugation, depending on the orientation of $\l$) for all unit cells. Because of translational invariance of the problem, the calculation reduces to a single link, e.g. ${\l_0}$: 
\begin{align}
   \mathcal{B} &= 4 \sum_{z^b_{{\l_0}}} \int_0^{2\pi n} d\theta^b_{{\v_1}}d\theta^b_{{\v_2}}\sum_{t^b_{{\v_1}},t^b_{{\v_2}}} \left(-\frac{pq}{n} \right) z^b_{\l_0}\bar{b}_{\l_0} \left(\frac{n}{2\pi}\right)^2 W^2 \mathrm{e}^{-\bar{b}_{\l_0}^2  (n/2\pi)^2 W^2} / \mathcal{N}' \ , \nonumber \\[.2cm]
    \bar{b}_{\l_0} &\equiv 2\pi  \left(z^b_{\l_0}+\frac{2\pi t^b_{{\v_1}} +\theta^b_{{\v_1}}-2\pi t^b_{{\v_2}} -\theta^b_{{\v_2}}}{2\pi n} \right)
\end{align}
where ${\v_1}$ and ${\v_2}$ are the two endpoints of link ${\l_0}$, $\mathcal{N}' = \frac{\sqrt{\pi}}{nW}\cdot (2\pi n)^2$ is a normalization factor, and we have neglected the infinitesimal background field. Then, the summation of $t^b$ and the integral over $\theta^b$ on $[0,2\pi)$ on each vertex can make up to an integral over $\theta^b$ on $[0,2\pi n)$:
\begin{align}
   \mathcal{B} &=  4\sum_{z^b_{{\l_0}}}  \int_0^{2\pi n} d\theta^b_{{\v_1}}d\theta^b_{{\v_2}} \left(-\frac{pq}{ n}\right)z^b_{\l_0}\bar{b}_{\l_0} \left(\frac{n}{2\pi}\right)^2W^2 \mathrm{e}^{-\bar{b}_{\l_0}^2  (n/2\pi)^2 W^2} / \mathcal{N}' \ , \nonumber \\[.2cm]
    \bar{b}_{\l_0} &\equiv 2\pi  \left(z^b_{\l_0}+\frac{ \theta^b_{{\v_1}}-\theta^b_{{\v_2}}}{2\pi n}\right) \ .
\end{align}
Integrating over $\theta^b_{{\v_2}}$ yields
\begin{align}
   \mathcal{B} &=  2\sum_{z^b_{{\l_0}}}  \int_0^{2\pi n} d\theta^b_{{\v_1}} (-pq)z^b_{\l_0} \left(\mathrm{e}^{-\left(z^b_{\l_0}-1+\frac{\theta^b_{{\v_1}}}{2\pi n}\right)^2 n^2W^2}-\mathrm{e}^{-\left(z^b_{\l_0}+\frac{\theta^b_{{\v_1}}}{2\pi n}\right)^2 n^2W^2}\right) / \mathcal{N}' \nonumber\\[.2cm]
    &=  (4\pi n) (-pq) \int_{-\infty}^{\infty} dz \ \mathrm{e}^{-z^2 n^2W^2} / \mathcal{N}' \ = \ -\frac{pq}{\pi n}
\end{align}
where in the second line we have used combined the summation over $z^b_{{\l_0}}$ and integral over $\theta^b_{{\v_1}}$ to make up an infinite integral. We thus find a constant Berry curvature all over the domain of integral. Substituting into \eqref{app:Chern} confirms that the Hall conductivity $2\pi \sigma_H = -2pq/n$.

In the calculation above, the only approximation we made is that the overlap between different Gaussian wave packets is negligible. The errors introduced by such overlaps are always exponentially suppressed by a factor of $\mathrm{e}^{-\pi^2/n^2W^2}$, and thus can be arbitrarily small in $W\sim \sqrt{\epsilon/U} \rightarrow 0$ limit. The errors introduced by using our trial wavefunctions rather than the exact ones, $\mathcal{O}(\sqrt{\epsilon/U}, \epsilon/V, V/\sqrt{\epsilon U})$, are also controlled to be small in the said limit. Since the integral \eqref{app:Chern} is the first Chern class of a $U(1)$ principal bundle of the ground state wave functions on the base manifold of a torus parametrized by the holonomies $\alpha_x$ and $\alpha_y$, it must be quantized. So the Hall conductivity must be the calculated fraction, as long as the excitation gap is not closed. This completes the proof of the exactness and robustness of the Hall conductivity.

\section{Identifying the Topological Orders for the Models in Ref.~\cite{geraedts2013exact}}
\label{app_relation}

At the end of Section \ref{sect_Hamiltonian} we briefly mentioned the relation between the present work and \cite{geraedts2013exact}. A family of Hamiltonians were introduced in \cite{geraedts2013exact}, with the ``$c=1$'' cases (in their notations) coincide with the Hamiltonians we constructed (their ``$d$'' corresponds to our $n$), although the motivating reasons are quite different. In a follow-up study \cite{geraedts2017lattice}, the relation between the ``$c=1$'' cases (in the absence of electromagnetic background) and the $\mathbb{Z}_n$ toric code was argued. On the other hand, the other cases of integer ``$c\neq 1$'' were more mysterious; the nature of their topological orders was not identified. Both the ``$c=1$'' and ``$c\neq 1$'' models were studied with sign-free Monte-Carlo numerical simulation; however, a connection to Chern-Simons theory was not established and solutions (or a path towards the solutions) were not obtained. In this appendix:
\begin{enumerate}
\item We first explain how to suitably identify the topological orders for the models considered in \cite{geraedts2013exact}, especially the mysterious ``$c\neq 1$'' cases. We will show that in the topological limit the ``$c\neq 1$'' models do not lead to new topological orders; they reduce to certain ``$c=1$'' cases. 
\item Then, based on this observation, we sketch how the solve the ``$c\neq 1$'' Hamiltonians in the topological limit. 
\end{enumerate}

In Appendix B.2 of \cite{geraedts2013exact} a mapping between their lattice models and Chern-Simons-like theories was attempted. However, even in the ``$c=1$'' cases which coincide with our Hamiltonians, the attempted mapping disagrees with our \eqref{spacetime_lattice_action}. More exactly, it was suggested in \cite{geraedts2013exact} that the lattice Hamiltonians map to Chern-Simons-like spacetime lattice theories similar to our \eqref{spacetime_lattice_action}, but with $1$ instead of $n$ in front of our second and third terms -- the Dirac string couplings. Here we emphasize that the Dirac string coupling coefficient being $n$ instead of $1$ is important, for it determines the topological order. To see this, for simplicity let us turn off the electromagnetic coupling, then, if we sum out the Dirac strings $s^a$ and $s^b$, the gauge fields $a, b$ are indeed restricted to $(2\pi/n)\mathbb{Z} \ \mod 2\pi$, as expected for a $\mathbb{Z}_n$ toric code. On the other hand, if the Dirac string coupling coefficient were $1$ instead of $n$, then $a, b$ would have been reduced to multiples of $2\pi$, but then the $a db$ term would always have dropped out for being a multiple of $2\pi$ and the theory would have become topologically trivial.

To suitably identify the topological order for the models in \cite{geraedts2013exact}, we should use their Eq. 20 instead. One may note that, upon rescaling their real valued $\alpha_1, \alpha_2$ by $d$ (our $n$) and taking the small $\lambda$ (our $\epsilon$) limit, the topological part of their Eq. 20 takes the form (in the below our integers $(m, n)$ correspond to $(c, d)$ in \cite{geraedts2013exact}):
\begin{align}
\frac{mn}{2\pi} \sum_{plaq. \ p} a_p (db)_p - n\sum_{plaq. \ p} a_p s^b_p - n\sum_{link \: l} s^a_l b_l + \sum_{cube \: c} \theta^a_c (ds^b)_c + \sum_{vert. \: v} \theta^b_v (\partial s^a)_v
\label{spacetime_lattice_action_modified}
\end{align}
which, when $m=1$, is our \eqref{spacetime_lattice_action} (dropping the electromagnetic coupling for now). This addresses the nature of the topological orders for the $m=1$ cases. 

How about the more mysterious $m\neq 1$ cases? To understand them, we first consider the level $n'=mn$ doubled Chern-Simons theory on lattice, i.e. \eqref{spacetime_lattice_action_modified} but with the Dirac string couplings being $n'=mn$ instead of $n$ -- this is simply the theory for $\mathbb{Z}_{n'}$ toric code. Then, what does it mean to reduce the Dirac string coupling from $n'=mn$ to $n$? This corresponds to condensing those Wilson loops $L^a$ and $L^b$ in \eqref{spacetime_lattice_action} when they take values (charges) $n\mathbb{Z} \mod n'$ under $a$ and $b$ respectively. Equivalently, this corresponds to gauging the 1-form $\mathbb{Z}_m\times \mathbb{Z}_m$ subgroup out of the global 1-form $\mathbb{Z}_{n'}\times \mathbb{Z}_{n'}$ symmetry \cite{Chen:2019mjw}. When $m$ does not divide $n$, this is in fact a classic example of gauging a 1-form symmetry with mixed anomaly. In general, when a theory contains gauge anomaly, some sectors of the anomalous theory will vanish, while some sectors will remain as a non-anomalous theory. Now we show that in the present case, the remaining non-anomalous theory is simply the theory of level $\wt{n}=n/\mathrm{gcd}(m, n)$.

In the present case, the anomaly is manifested by the fact that the Wilson loops that we are trying to condense have a mutual statistics of
\begin{align}
\exp\left[\i\frac{2\pi}{n'} (n l^a) (n l^b)\right] = \exp\left[\i\frac{2\pi n}{m} l^a l^b\right]
\end{align}
where $nl^a \mod n'$ is the $a$ charge of some Wilson loop $L^a$, and $nl^b \mod n'$ is the $b$ charge of some $L^b$ that has linking number $1$ with $L^a$. When $m$ divides $n$, the mutual statistics is trivial and the Wilson loops can indeed be simultaneously condensed. In this case we arrive at a level $\wt{n} = n/m$ theory -- this can be seen by simply rescaling the real valued $a, b$ in \eqref{spacetime_lattice_action_modified} by $1/m$. In the opposite scenario where $m, n$ are coprime and $m\neq 1$, for fixed $l^a$, summing over all possible $l^b$ would yield a cancelled contribution to the partition function unless $l^a=0 \mod m$, i.e. $nl^a = 0 \mod n'$. This is equivalent to replacing $s^b\rightarrow m s^b$, i.e. restoring the coefficient $n'=mn$ in front of the $a s^b$ term in \eqref{spacetime_lattice_action_modified}, while still keeping $n$ in front of the $s^a b$ term. This theory, however, is nothing but a level $n$ theory -- as can be seen from rescaling the real valued $a$ by $1/m$. (The ``particle-hole dual'' $m=n-1$ case mentioned in \cite{geraedts2017lattice} is a special case of this scenario.) Now we are ready to explain the situation for general values of $m$ and $n$. Define the integers $\wt{m}, \wt{n}$ to be the ones such that $m/\wt{m}=n/\wt{n}=\mathrm{gcd}(m, n)$, the greatest common divisor of $m$ and $n$. For fixed $l^a$, summing over all possible $l^b$ would yield a cancelled contribution to the partition function unless $l^a=0 \mod \wt{m}$. But this is equivalent to replacing $s^b\rightarrow \wt{m} s^b$, i.e.
\begin{align}
\frac{mn}{2\pi} \sum_{plaq. \ p} a_p (db)_p - \wt{m} n\sum_{plaq. \ p} a_p s^b_p - n\sum_{link \: l} s^a_l b_l + \sum_{cube \: c} \theta^a_c (ds^b)_c + \sum_{vert. \: v} \theta^b_v (\partial s^a)_v.
\end{align}
Now we may rescale the real valued $a$ and $b$ by $1/m$ and $\wt{n}/n$ respectively, and we find the theory in fact reduces to a level $\wt{n}$ theory. This completes the identification of the topological orders for the models in \cite{geraedts2013exact} -- the general values of ``$c$'' (our $m$) are topologically equivalent to ``$c=1$'' but with a changed value of ``$d$'' (our $n$).

In the above we only considered the intrinsic topological order. But then the coupling to electromagnetic background is straightforward. If the $s^a$ and $s^b$ in the original anomalous theory \eqref{spacetime_lattice_action_modified} are coupled to $A$ and $A^\star$ with coefficients $q$ and $p$ respectively, then in the remaining non-anomalous level $\wt{n}$ theory, the couplings are $\wt{q}=q$ and $\wt{p}=\wt{m} p$ respectively. The Hall conductivity is then $-2\wt{m} pq/\wt{n} = -2m pq/n$, in agreement with the numerically computed current-current correlation from \cite{geraedts2013exact, geraedts2017lattice} when setting $q=p=1$.

Now that we have understood the topological nature of the Hamiltonians for $m\neq 1$ (``i.e. $c\neq 1$'' in \cite{geraedts2013exact}), we show that these Hamiltonians, which have the integer $n$ replaced by the fraction $n/m$ in both the commutation relations \eqref{ba_comm} and the Hamiltonian $H$, can still be controllably solved (applying the bounding function $F$ to the $V$ terms in understood) by our method with only slight modification. A potential problem is that the $\cos((n/m)a_{\l^\star})$ and $\cos((n/m)b_{\l})$ in the Hamiltonian no longer commute (now $[b_{\l}, a_{\l^\star}]=\i 2\pi m/n$) if $m$ does not divide $n$. In the solving procedure, on each link we first solve $H_{\mathrm{link}, b}$ and find the narrow Gaussians (in the $b$ basis) peaked around $\bar{b}_j =(2\pi m/n) j + \bar{b}_0, \ j\in\mathbb{Z}$. The non-commutativity issues is manifested by the fact that the set of such narrow Gaussians with  $j\in\mathbb{Z}$ no longer form an invariant subspace when we consider the shift terms $e^{\pm \i (n/m)a}$ in $H_{\mathrm{link}, a}$, which shift a narrow Gaussian by a distance of $2\pi$. However, this is not a substantial problem. It is not hard to see that the only consequence at low energy is that the low energy states are then the ones formed by the peaks that are commensurate with each other. That is, the shifts only matter at low energy when they act for $\wt{m}$ times, i.e. $(e^{\pm \i (n/m)a})^{\wt{m}}$, shifting a narrow Gaussian peak to another one located $2\pi\wt{m}$ away, separated by $\wt{n}-1$ other peaks in between. Thus, the only change is that effective period for the low energy states is $2\pi\wt{m}$ instead of $2\pi$, and there are $\wt{n}$ low energy states on each link, as expected based on the previous identification of topological order. Corresponding, since we are only keeping $(e^{\pm \i (n/m)a})^{\wt{m}}$, the coupling of $e^{\pm \i (n/m)a}$ to $e^{\mp\i pA^\star}$ means $p$ is effectively rescaled to $\wt{m}p$. The remaining parts of the solution are unchanged. Hence we still can controllably solve the Hamiltonian, and the solution, as expected from the reasoning before, is essentially the same as that for a level $\wt{n}$ theory, with electromagnetic couplings $q$ and $\wt{m} p$.

\end{widetext}

\bibliography{FQH_Lattice}

\begin{thebibliography}{24}%
\makeatletter
\providecommand \@ifxundefined [1]{%
 \@ifx{#1\undefined}
}%
\providecommand \@ifnum [1]{%
 \ifnum #1\expandafter \@firstoftwo
 \else \expandafter \@secondoftwo
 \fi
}%
\providecommand \@ifx [1]{%
 \ifx #1\expandafter \@firstoftwo
 \else \expandafter \@secondoftwo
 \fi
}%
\providecommand \natexlab [1]{#1}%
\providecommand \enquote  [1]{``#1''}%
\providecommand \bibnamefont  [1]{#1}%
\providecommand \bibfnamefont [1]{#1}%
\providecommand \citenamefont [1]{#1}%
\providecommand \href@noop [0]{\@secondoftwo}%
\providecommand \href [0]{\begingroup \@sanitize@url \@href}%
\providecommand \@href[1]{\@@startlink{#1}\@@href}%
\providecommand \@@href[1]{\endgroup#1\@@endlink}%
\providecommand \@sanitize@url [0]{\catcode `\\12\catcode `\$12\catcode
  `\&12\catcode `\#12\catcode `\^12\catcode `\_12\catcode `\%12\relax}%
\providecommand \@@startlink[1]{}%
\providecommand \@@endlink[0]{}%
\providecommand \url  [0]{\begingroup\@sanitize@url \@url }%
\providecommand \@url [1]{\endgroup\@href {#1}{\urlprefix }}%
\providecommand \urlprefix  [0]{URL }%
\providecommand \Eprint [0]{\href }%
\providecommand \doibase [0]{http://dx.doi.org/}%
\providecommand \selectlanguage [0]{\@gobble}%
\providecommand \bibinfo  [0]{\@secondoftwo}%
\providecommand \bibfield  [0]{\@secondoftwo}%
\providecommand \translation [1]{[#1]}%
\providecommand \BibitemOpen [0]{}%
\providecommand \bibitemStop [0]{}%
\providecommand \bibitemNoStop [0]{.\EOS\space}%
\providecommand \EOS [0]{\spacefactor3000\relax}%
\providecommand \BibitemShut  [1]{\csname bibitem#1\endcsname}%
\let\auto@bib@innerbib\@empty
\bibitem [{\citenamefont {Klitzing}\ \emph {et~al.}(1980)\citenamefont
  {Klitzing}, \citenamefont {Dorda},\ and\ \citenamefont
  {Pepper}}]{klitzing1980new}%
  \BibitemOpen
  \bibfield  {author} {\bibinfo {author} {\bibfnamefont {K.~v.}\ \bibnamefont
  {Klitzing}}, \bibinfo {author} {\bibfnamefont {G.}~\bibnamefont {Dorda}}, \
  and\ \bibinfo {author} {\bibfnamefont {M.}~\bibnamefont {Pepper}},\ }\href
  {\doibase doi.org/10.1103/PhysRevLett.45.494} {\bibfield  {journal} {\bibinfo
   {journal} {Physical review letters}\ }\textbf {\bibinfo {volume} {45}},\
  \bibinfo {pages} {494} (\bibinfo {year} {1980})}\BibitemShut {NoStop}%
\bibitem [{\citenamefont {Tsui}\ \emph {et~al.}(1982)\citenamefont {Tsui},
  \citenamefont {Stormer},\ and\ \citenamefont {Gossard}}]{tsui1982two}%
  \BibitemOpen
  \bibfield  {author} {\bibinfo {author} {\bibfnamefont {D.~C.}\ \bibnamefont
  {Tsui}}, \bibinfo {author} {\bibfnamefont {H.~L.}\ \bibnamefont {Stormer}}, \
  and\ \bibinfo {author} {\bibfnamefont {A.~C.}\ \bibnamefont {Gossard}},\
  }\href {\doibase 10.1103/PhysRevLett.48.1559} {\bibfield  {journal} {\bibinfo
   {journal} {Physical Review Letters}\ }\textbf {\bibinfo {volume} {48}},\
  \bibinfo {pages} {1559} (\bibinfo {year} {1982})}\BibitemShut {NoStop}%
\bibitem [{\citenamefont {Laughlin}(1983)}]{Laughlin:1983fy}%
  \BibitemOpen
  \bibfield  {author} {\bibinfo {author} {\bibfnamefont {R.~B.}\ \bibnamefont
  {Laughlin}},\ }\href {\doibase 10.1103/PhysRevLett.50.1395} {\bibfield
  {journal} {\bibinfo  {journal} {Phys. Rev. Lett.}\ }\textbf {\bibinfo
  {volume} {50}},\ \bibinfo {pages} {1395} (\bibinfo {year} {1983})},\ \bibinfo
  {note} {[,308(1983)]}\BibitemShut {NoStop}%
\bibitem [{\citenamefont {Kitaev}(2003)}]{Kitaev:1997wr}%
  \BibitemOpen
  \bibfield  {author} {\bibinfo {author} {\bibfnamefont {A.}~\bibnamefont
  {Kitaev}},\ }\href {\doibase 10.1016/S0003-4916(02)00018-0} {\bibfield
  {journal} {\bibinfo  {journal} {Annals Phys.}\ }\textbf {\bibinfo {volume}
  {303}},\ \bibinfo {pages} {2} (\bibinfo {year} {2003})},\ \Eprint
  {http://arxiv.org/abs/quant-ph/9707021} {arXiv:quant-ph/9707021 [quant-ph]}
  \BibitemShut {NoStop}%
\bibitem [{\citenamefont {Levin}\ and\ \citenamefont
  {Wen}(2005)}]{Levin:2004mi}%
  \BibitemOpen
  \bibfield  {author} {\bibinfo {author} {\bibfnamefont {M.~A.}\ \bibnamefont
  {Levin}}\ and\ \bibinfo {author} {\bibfnamefont {X.-G.}\ \bibnamefont
  {Wen}},\ }\href {\doibase 10.1103/PhysRevB.71.045110} {\bibfield  {journal}
  {\bibinfo  {journal} {Phys. Rev.}\ }\textbf {\bibinfo {volume} {B71}},\
  \bibinfo {pages} {045110} (\bibinfo {year} {2005})},\ \Eprint
  {http://arxiv.org/abs/cond-mat/0404617} {arXiv:cond-mat/0404617
  [cond-mat.str-el]} \BibitemShut {NoStop}%
\bibitem [{\citenamefont {Chen}\ \emph {et~al.}(2013)\citenamefont {Chen},
  \citenamefont {Gu}, \citenamefont {Liu},\ and\ \citenamefont
  {Wen}}]{Chen:2011pg}%
  \BibitemOpen
  \bibfield  {author} {\bibinfo {author} {\bibfnamefont {X.}~\bibnamefont
  {Chen}}, \bibinfo {author} {\bibfnamefont {Z.-C.}\ \bibnamefont {Gu}},
  \bibinfo {author} {\bibfnamefont {Z.-X.}\ \bibnamefont {Liu}}, \ and\
  \bibinfo {author} {\bibfnamefont {X.-G.}\ \bibnamefont {Wen}},\ }\href
  {\doibase 10.1103/PhysRevB.87.155114} {\bibfield  {journal} {\bibinfo
  {journal} {Phys. Rev.}\ }\textbf {\bibinfo {volume} {B87}},\ \bibinfo {pages}
  {155114} (\bibinfo {year} {2013})},\ \Eprint {http://arxiv.org/abs/1106.4772}
  {arXiv:1106.4772 [cond-mat.str-el]} \BibitemShut {NoStop}%
\bibitem [{\citenamefont {Turaev}\ and\ \citenamefont
  {Viro}(1992)}]{Turaev:1992hq}%
  \BibitemOpen
  \bibfield  {author} {\bibinfo {author} {\bibfnamefont {V.~G.}\ \bibnamefont
  {Turaev}}\ and\ \bibinfo {author} {\bibfnamefont {O.~Y.}\ \bibnamefont
  {Viro}},\ }\href {\doibase 10.1016/0040-9383(92)90015-A} {\bibfield
  {journal} {\bibinfo  {journal} {Topology}\ }\textbf {\bibinfo {volume}
  {31}},\ \bibinfo {pages} {865} (\bibinfo {year} {1992})}\BibitemShut
  {NoStop}%
\bibitem [{\citenamefont {Kirillov}(2011)}]{Kirillov:2011mk}%
  \BibitemOpen
  \bibfield  {author} {\bibinfo {author} {\bibfnamefont {A.}~\bibnamefont
  {Kirillov}, \bibfnamefont {Jr}},\ }\href@noop {} {\  (\bibinfo {year}
  {2011})},\ \Eprint {http://arxiv.org/abs/1106.6033} {arXiv:1106.6033
  [math.AT]} \BibitemShut {NoStop}%
\bibitem [{\citenamefont {Kitaev}\ and\ \citenamefont
  {Kong}(2012)}]{Kitaev:2011dxc}%
  \BibitemOpen
  \bibfield  {author} {\bibinfo {author} {\bibfnamefont {A.}~\bibnamefont
  {Kitaev}}\ and\ \bibinfo {author} {\bibfnamefont {L.}~\bibnamefont {Kong}},\
  }\href {\doibase 10.1007/s00220-012-1500-5} {\bibfield  {journal} {\bibinfo
  {journal} {Commun. Math. Phys.}\ }\textbf {\bibinfo {volume} {313}},\
  \bibinfo {pages} {351} (\bibinfo {year} {2012})},\ \Eprint
  {http://arxiv.org/abs/1104.5047} {arXiv:1104.5047 [cond-mat.str-el]}
  \BibitemShut {NoStop}%
\bibitem [{\citenamefont {Kapustin}\ and\ \citenamefont
  {Fidkowski}(2020)}]{kapustin2020local}%
  \BibitemOpen
  \bibfield  {author} {\bibinfo {author} {\bibfnamefont {A.}~\bibnamefont
  {Kapustin}}\ and\ \bibinfo {author} {\bibfnamefont {L.}~\bibnamefont
  {Fidkowski}},\ }\href {\doibase 10.1007/s00220-019-03444-1} {\bibfield
  {journal} {\bibinfo  {journal} {Commun. Math. Phys.}\ }\textbf {\bibinfo
  {volume} {373}},\ \bibinfo {pages} {763} (\bibinfo {year} {2020})},\ \Eprint
  {http://arxiv.org/abs/1810.07756} {arXiv:1810.07756 [cond-mat.str-el]}
  \BibitemShut {NoStop}%
\bibitem [{\citenamefont {Chen}(2021)}]{Chen:2019mjw}%
  \BibitemOpen
  \bibfield  {author} {\bibinfo {author} {\bibfnamefont {J.-Y.}\ \bibnamefont
  {Chen}},\ }\href {\doibase 10.1007/s00220-020-03927-6} {\bibfield  {journal}
  {\bibinfo  {journal} {Commun. Math. Phys.}\ }\textbf {\bibinfo {volume}
  {381}},\ \bibinfo {pages} {293} (\bibinfo {year} {2021})},\ \Eprint
  {http://arxiv.org/abs/1902.06756} {arXiv:1902.06756 [cond-mat.str-el]}
  \BibitemShut {NoStop}%
\bibitem [{\citenamefont {Kapustin}\ and\ \citenamefont
  {Saulina}(2011)}]{Kapustin:2010hk}%
  \BibitemOpen
  \bibfield  {author} {\bibinfo {author} {\bibfnamefont {A.}~\bibnamefont
  {Kapustin}}\ and\ \bibinfo {author} {\bibfnamefont {N.}~\bibnamefont
  {Saulina}},\ }\href {\doibase 10.1016/j.nuclphysb.2010.12.017} {\bibfield
  {journal} {\bibinfo  {journal} {Nucl. Phys.}\ }\textbf {\bibinfo {volume}
  {B845}},\ \bibinfo {pages} {393} (\bibinfo {year} {2011})},\ \Eprint
  {http://arxiv.org/abs/1008.0654} {arXiv:1008.0654 [hep-th]} \BibitemShut
  {NoStop}%
\bibitem [{\citenamefont {Lin}\ and\ \citenamefont
  {Levin}(2014)}]{Lin:2014aca}%
  \BibitemOpen
  \bibfield  {author} {\bibinfo {author} {\bibfnamefont {C.-H.}\ \bibnamefont
  {Lin}}\ and\ \bibinfo {author} {\bibfnamefont {M.}~\bibnamefont {Levin}},\
  }\href {\doibase 10.1103/PhysRevB.89.195130} {\bibfield  {journal} {\bibinfo
  {journal} {Phys. Rev.}\ }\textbf {\bibinfo {volume} {B89}},\ \bibinfo {pages}
  {195130} (\bibinfo {year} {2014})},\ \Eprint {http://arxiv.org/abs/1402.4081}
  {arXiv:1402.4081 [cond-mat.str-el]} \BibitemShut {NoStop}%
\bibitem [{\citenamefont {Kitaev}(2006)}]{Kitaev:2006lla}%
  \BibitemOpen
  \bibfield  {author} {\bibinfo {author} {\bibfnamefont {A.}~\bibnamefont
  {Kitaev}},\ }\href {\doibase 10.1016/j.aop.2005.10.005} {\bibfield  {journal}
  {\bibinfo  {journal} {Annals Phys.}\ }\textbf {\bibinfo {volume} {321}},\
  \bibinfo {pages} {2} (\bibinfo {year} {2006})},\ \Eprint
  {http://arxiv.org/abs/cond-mat/0506438} {arXiv:cond-mat/0506438
  [cond-mat.mes-hall]} \BibitemShut {NoStop}%
\bibitem [{\citenamefont {Geraedts}\ and\ \citenamefont
  {Motrunich}(2013)}]{geraedts2013exact}%
  \BibitemOpen
  \bibfield  {author} {\bibinfo {author} {\bibfnamefont {S.~D.}\ \bibnamefont
  {Geraedts}}\ and\ \bibinfo {author} {\bibfnamefont {O.~I.}\ \bibnamefont
  {Motrunich}},\ }\href {\doibase 10.1016/j.aop.2013.03.017} {\bibfield
  {journal} {\bibinfo  {journal} {Annals of Physics}\ }\textbf {\bibinfo
  {volume} {334}},\ \bibinfo {pages} {288} (\bibinfo {year} {2013})},\ \Eprint
  {http://arxiv.org/abs/1302.1436} {arXiv:1302.1436 [cond-mat.str-el]}
  \BibitemShut {NoStop}%
\bibitem [{\citenamefont {Geraedts}\ and\ \citenamefont
  {Motrunich}(2017)}]{geraedts2017lattice}%
  \BibitemOpen
  \bibfield  {author} {\bibinfo {author} {\bibfnamefont {S.~D.}\ \bibnamefont
  {Geraedts}}\ and\ \bibinfo {author} {\bibfnamefont {O.~I.}\ \bibnamefont
  {Motrunich}},\ }\href {\doibase 10.1103/PhysRevB.96.115137} {\bibfield
  {journal} {\bibinfo  {journal} {Physical Review B}\ }\textbf {\bibinfo
  {volume} {96}},\ \bibinfo {pages} {115137} (\bibinfo {year} {2017})},\
  \Eprint {http://arxiv.org/abs/1705.06308} {arXiv:1705.06308
  [cond-mat.str-el]} \BibitemShut {NoStop}%
\bibitem [{\citenamefont {DeMarco}\ and\ \citenamefont
  {Wen}(2021)}]{DeMarco:2021erp}%
  \BibitemOpen
  \bibfield  {author} {\bibinfo {author} {\bibfnamefont {M.}~\bibnamefont
  {DeMarco}}\ and\ \bibinfo {author} {\bibfnamefont {X.-G.}\ \bibnamefont
  {Wen}},\ }\href@noop {} {\  (\bibinfo {year} {2021})},\ \Eprint
  {http://arxiv.org/abs/2102.13057} {arXiv:2102.13057 [cond-mat.str-el]}
  \BibitemShut {NoStop}%
\bibitem [{\citenamefont {Witten}(2003)}]{Witten:2003ya}%
  \BibitemOpen
  \bibfield  {author} {\bibinfo {author} {\bibfnamefont {E.}~\bibnamefont
  {Witten}},\ }\href@noop {} {\  (\bibinfo {year} {2003})},\ \Eprint
  {http://arxiv.org/abs/hep-th/0307041} {arXiv:hep-th/0307041 [hep-th]}
  \BibitemShut {NoStop}%
\bibitem [{\citenamefont {Wen}(1995)}]{Wen:1995qn}%
  \BibitemOpen
  \bibfield  {author} {\bibinfo {author} {\bibfnamefont {X.-G.}\ \bibnamefont
  {Wen}},\ }\href {\doibase 10.1080/00018739500101566} {\bibfield  {journal}
  {\bibinfo  {journal} {Adv. Phys.}\ }\textbf {\bibinfo {volume} {44}},\
  \bibinfo {pages} {405} (\bibinfo {year} {1995})},\ \Eprint
  {http://arxiv.org/abs/cond-mat/9506066} {arXiv:cond-mat/9506066 [cond-mat]}
  \BibitemShut {NoStop}%
\bibitem [{\citenamefont {Gaiotto}\ \emph {et~al.}(2015)\citenamefont
  {Gaiotto}, \citenamefont {Kapustin}, \citenamefont {Seiberg},\ and\
  \citenamefont {Willett}}]{Gaiotto:2014kfa}%
  \BibitemOpen
  \bibfield  {author} {\bibinfo {author} {\bibfnamefont {D.}~\bibnamefont
  {Gaiotto}}, \bibinfo {author} {\bibfnamefont {A.}~\bibnamefont {Kapustin}},
  \bibinfo {author} {\bibfnamefont {N.}~\bibnamefont {Seiberg}}, \ and\
  \bibinfo {author} {\bibfnamefont {B.}~\bibnamefont {Willett}},\ }\href
  {\doibase 10.1007/JHEP02(2015)172} {\bibfield  {journal} {\bibinfo  {journal}
  {JHEP}\ }\textbf {\bibinfo {volume} {02}},\ \bibinfo {pages} {172} (\bibinfo
  {year} {2015})},\ \Eprint {http://arxiv.org/abs/1412.5148} {arXiv:1412.5148
  [hep-th]} \BibitemShut {NoStop}%
\bibitem [{\citenamefont {Levin}\ \emph {et~al.}(2011)\citenamefont {Levin},
  \citenamefont {Burnell}, \citenamefont {Koch-Janusz},\ and\ \citenamefont
  {Stern}}]{Levin:2011hq}%
  \BibitemOpen
  \bibfield  {author} {\bibinfo {author} {\bibfnamefont {M.}~\bibnamefont
  {Levin}}, \bibinfo {author} {\bibfnamefont {F.~J.}\ \bibnamefont {Burnell}},
  \bibinfo {author} {\bibfnamefont {M.}~\bibnamefont {Koch-Janusz}}, \ and\
  \bibinfo {author} {\bibfnamefont {A.}~\bibnamefont {Stern}},\ }\href
  {\doibase 10.1103/PhysRevB.84.235145} {\bibfield  {journal} {\bibinfo
  {journal} {Phys. Rev.}\ }\textbf {\bibinfo {volume} {B84}},\ \bibinfo {pages}
  {235145} (\bibinfo {year} {2011})},\ \Eprint {http://arxiv.org/abs/1108.4954}
  {arXiv:1108.4954 [cond-mat.str-el]} \BibitemShut {NoStop}%
\bibitem [{\citenamefont {Niu}\ \emph {et~al.}(1985)\citenamefont {Niu},
  \citenamefont {Thouless},\ and\ \citenamefont {Wu}}]{niu1985quantized}%
  \BibitemOpen
  \bibfield  {author} {\bibinfo {author} {\bibfnamefont {Q.}~\bibnamefont
  {Niu}}, \bibinfo {author} {\bibfnamefont {D.~J.}\ \bibnamefont {Thouless}}, \
  and\ \bibinfo {author} {\bibfnamefont {Y.-S.}\ \bibnamefont {Wu}},\ }\href
  {\doibase 10.1103/PhysRevB.31.3372} {\bibfield  {journal} {\bibinfo
  {journal} {Physical Review B}\ }\textbf {\bibinfo {volume} {31}},\ \bibinfo
  {pages} {3372} (\bibinfo {year} {1985})}\BibitemShut {NoStop}%
\bibitem [{\citenamefont {Bezrukavnikov}\ and\ \citenamefont
  {Kapustin}(2019)}]{bezrukavnikov2019localization}%
  \BibitemOpen
  \bibfield  {author} {\bibinfo {author} {\bibfnamefont {R.}~\bibnamefont
  {Bezrukavnikov}}\ and\ \bibinfo {author} {\bibfnamefont {A.}~\bibnamefont
  {Kapustin}},\ }\href {\doibase 10.1007/s40598-019-00098-8} {\bibfield
  {journal} {\bibinfo  {journal} {Arnold Mathematical Journal}\ }\textbf
  {\bibinfo {volume} {5}},\ \bibinfo {pages} {15} (\bibinfo {year} {2019})},\
  \Eprint {http://arxiv.org/abs/1808.07602} {arXiv:1808.07602
  [cond-mat.mes-hall]} \BibitemShut {NoStop}%
\bibitem [{\citenamefont {Wang}\ and\ \citenamefont
  {Cheng}(2021)}]{Wang:2021smv}%
  \BibitemOpen
  \bibfield  {author} {\bibinfo {author} {\bibfnamefont {Q.-R.}\ \bibnamefont
  {Wang}}\ and\ \bibinfo {author} {\bibfnamefont {M.}~\bibnamefont {Cheng}},\
  }\href@noop {} {\  (\bibinfo {year} {2021})},\ \Eprint
  {http://arxiv.org/abs/2103.13399} {arXiv:2103.13399 [cond-mat.str-el]}
  \BibitemShut {NoStop}%
\end{thebibliography}%
\bibliographystyle{apsrev4-1} 

\end{document}